\documentclass[sigconf, screen]{acmart}
\usepackage{multirow}
\usepackage{caption}
\usepackage{subcaption}
\usepackage{algorithm}
\usepackage{algpseudocode}
\usepackage{colortbl}
\usepackage{hyperref}
\usepackage{tabularx}   
\makeatletter
\renewcommand{\algorithmiccomment}[1]{\bgroup\textit{// #1}\egroup}
\makeatother

\AtBeginDocument{%
  }

\copyrightyear{2025}
\acmYear{2025}
\setcopyright{acmlicensed}
\acmConference[MM '25] {Proceedings of the 33rd ACM International Conference on Multimedia}{October 27--31, 2025}{Dublin, Ireland.}
\acmBooktitle{Proceedings of the 33rd ACM International Conference on Multimedia (MM '25), October 27--31, 2025, Dublin, Ireland}
\acmISBN{979-8-4007-2035-2/2025/10}
\acmDOI{10.1145/3746027.3754751}

\begin{document}

\title{To Remember, To Adapt, To Preempt: A Stable Continual Test-Time Adaptation Framework for Remote Physiological Measurement in Dynamic Domain Shifts}
\renewcommand{\shorttitle}{PhysRAP: A Stable Continual Test-Time Adaptation Framework for Remote Physiological Measurement}


\author{Shuyang Chu}
\email{xjtucsy@stu.xjtu.edu.cn}
\affiliation{%
  \institution{School of Software Engineering\\
  Xi’an Jiaotong University}
  \city{Xi'an}
  \country{China}
}

\author{Jingang Shi}
\authornote{Corresponding author: Jingang Shi}
\email{jingang@xjtu.edu.cn}
\affiliation{%
  \institution{School of Software Engineering\\
  Xi'an Jiaotong University}
  \city{Xi'an}
  \country{China}}

\author{Xu Cheng}
\email{xcheng@nuist.edu.cn}
\affiliation{%
  \institution{School of Computer Science\\
  Nanjing University of Information Science and Technology}
  \city{Nanjing}
  \country{China}}

\author{Haoyu Chen}
\email{chen.haoyu@oulu.fi}
\affiliation{%
  \institution{CMVS\\
  University of Oulu}
  \city{Oulu}
  \country{Finland}}

\author{Xin Liu}
\email{linuxsino@gmail.com}
\affiliation{%
  \institution{School of Electrical and Information Engineering\\Tianjin University}
  \city{Tianjin}
  \country{China}}

\author{Jian Xu}
\email{xujian_xiyou@126.com}
\affiliation{%
  \institution{Shaanxi Key Laboratory of Information Communication Network and Security\\
  Xi'an University of Posts \& Telecommunications}
  \city{Xi'an}
  \country{China}}

\author{Guoying Zhao}
\email{guoying.zhao@oulu.fi}
\affiliation{%
  \institution{CMVS\\
  University of Oulu}
  \city{Oulu}
  \country{Finland}}






\renewcommand{\shortauthors}{Shuyang Chu et al.}

\begin{abstract}
Remote photoplethysmography (rPPG) aims to extract non-contact physiological signals from facial videos, which has recently shown great potential. Although existing rPPG approaches are making progress, they struggle to bridge the gap between source and target domains. Recent test-time adaptation (TTA) solutions typically optimize rPPG model for the incoming test videos using self-training loss under an unrealistic assumption that the target domain remains stationary. However, time-varying factors such as weather and lighting in dynamic environments often lead to continual domain shifts. The accumulation of erroneous gradients resulting from these shifts may corrupt the model's key parameters for identifying physiological information, leading to catastrophic forgetting. To retain the physiology-related knowledge in dynamic environments, we propose a physiology-related parameters freezing strategy. This strategy isolates physiology-related and domain-related parameters by assessing the model's uncertainty to current domain. It then freezes the physiology-related parameters during the adaptation process to prevent catastrophic forgetting. Moreover, the dynamic domain shifts typically display various characteristics in non-physiological information. It may lead to conflicting optimization objectives among domains during the TTA process, which is manifested as the over-adapted model losing its ability to adapt to future domains. To address over-adaptation, we propose a preemptive gradient modification strategy. This strategy preemptively adapts to potential future domains and uses the obtained gradients to modify the current adaptation, thereby preserving the model's adaptability in dynamic domain shifts. In summary, this paper proposes a stable continual test-time adaptation (CTTA) framework for rPPG measurement. We envision that the framework should \textbf{{R}}emember the past, \textbf{{A}}dapt to the present, and {\textbf{P}}reempt the future, denoted as \textbf{PhysRAP}. Extensive experiments show that our method achieves state-of-the-art performance, especially in continual domain shifts. The code is available at \href{https://github.com/xjtucsy/PhysRAP}{https://github.com/xjtucsy/PhysRAP}.
\end{abstract}



\begin{teaserfigure}
    \vspace{-1.5em}
  \includegraphics[width=\textwidth]{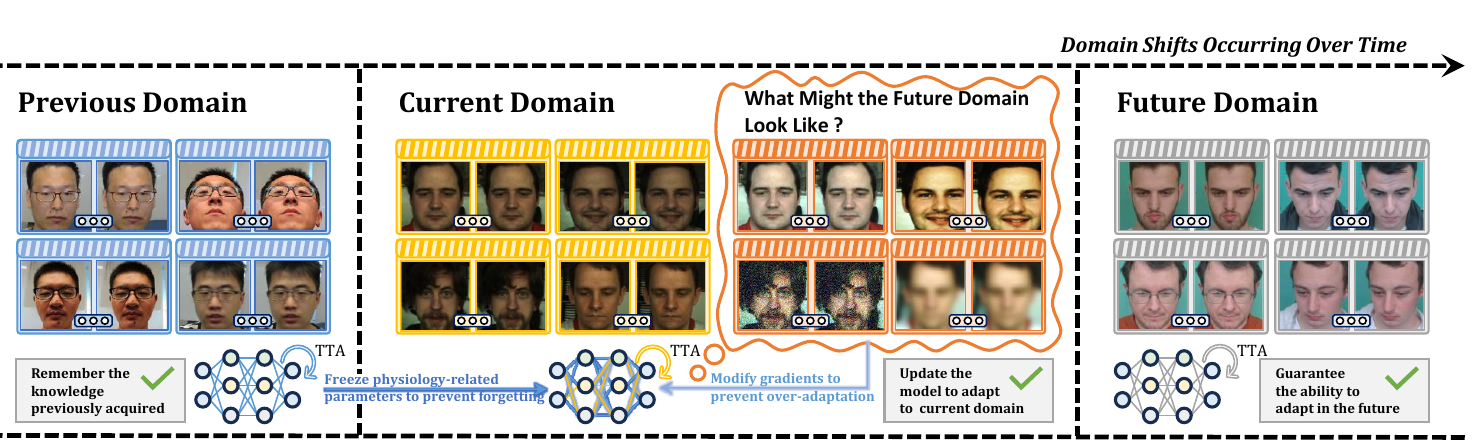}
  \caption{Visualization of the adaptation procedures of PhysRAP in dynamic domain shifts. Unlike standard TTA, which updates the entire model to fit the current domain, PhysRAP has two distinctive aspects. First, it identifies and freezes physiology-related parameters, achieving adaptation with minimal updates to prevent catastrophic forgetting. Second, PhysRAP preemptively considers potential future domain distributions, dynamically adjusting the updating strategy to avoid over-adaptation. PhysRAP provides stable and accurate rPPG measurement even when domains continually shift.}
  \label{fig:teaser}
  \Description{..}
\end{teaserfigure}


\maketitle
\vspace{-0.4em}
\section{Introduction}
\label{sec:intro}
\begin{figure}[t]
\centering
\includegraphics[width=\columnwidth]{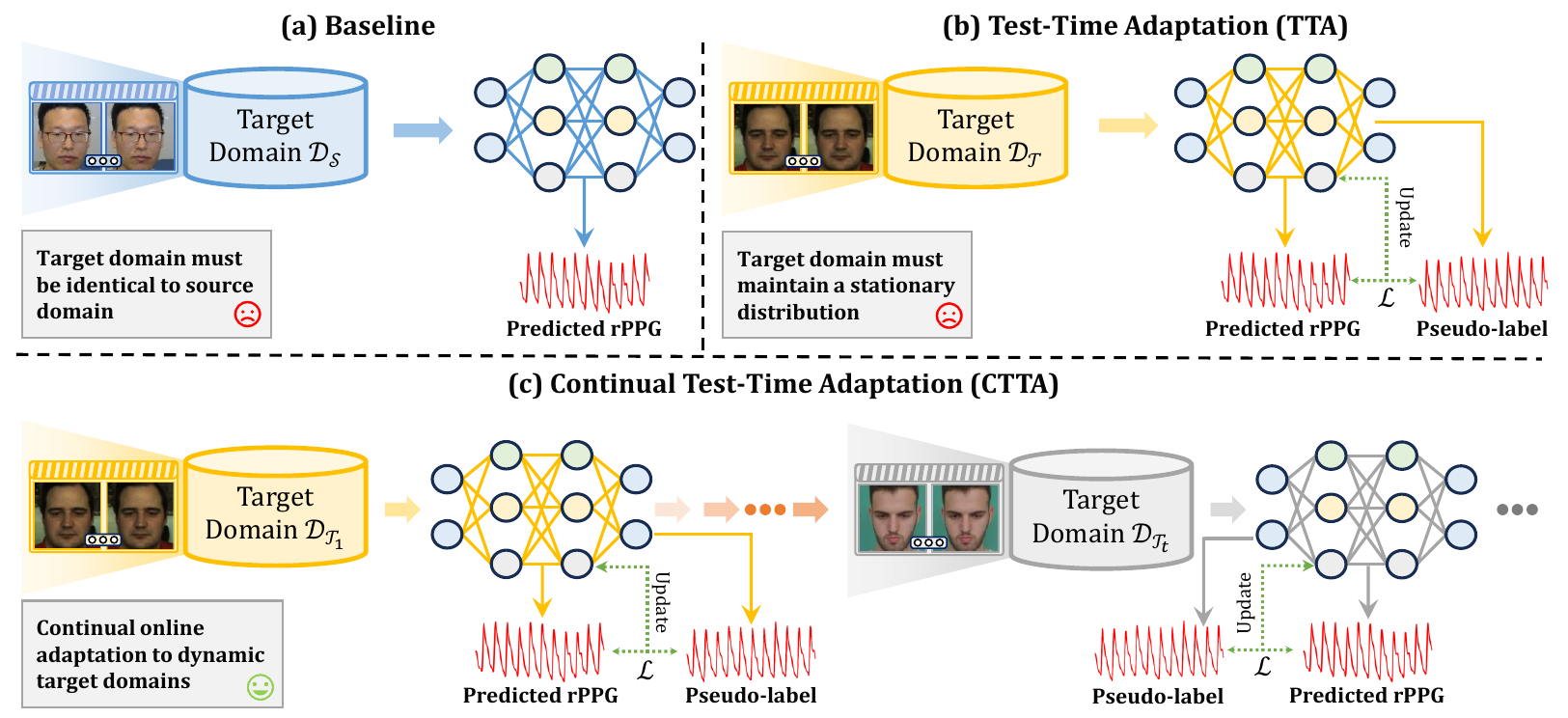}
\caption{Visualization of the testing phase of baseline, Test-Time Adaptation (TTA), and Continual Test-Time Adaptation (CTTA).}
\label{fig:comparsion_abc}\vspace{-1em}
\Description{..}
\end{figure}
Heart rate (HR) reflects the health status of the human body, which is widely used as a key indicator for physiological health monitoring. Currently, most clinical applications track heart activities by electrocardiography (ECG) and photoplethysmography (PPG). However, the complex setup and limited scalability make it difficult to use in real-world scenarios \cite{NiuRhythm2020}. To address this limitation, remote photoplethysmography (rPPG) aims to estimate heart rate from facial videos, which has recently gained increasing attention \cite{YuPhysNet2019,NiuCVD2020,YuPhysFormer2022,LuNEST2023,HiBaLu2024}. Besides, rPPG measurement has been used in other scenarios such as information security \cite{ChenPulseEdit2022} and telehealth \cite{LiuTSCAN2020}.

Early traditional methods \cite{PohICA2010,DeCHROM2013,DeI2014} often rely on blind signal decomposition and color space transformation, the effectiveness of which could be guaranteed only under strong assumptions. To make rPPG measurement more applicable to real-world scenarios, researchers have developed various deep learning-based approaches \cite{YuPhysNet2019,YuPhysFormer2022,YuPhysformer++2023,LiuEfficientPhys2021,ClusterPhysWei2024,liu2020Deeprppg}, which typically assume that the distribution of training and test data is consistent, as shown in Fig. \ref{fig:comparsion_abc}a. However, due to the variation of scenarios in real applications, the target domain distribution often differs from the source domain and changes continually (e.g., time-varying environmental conditions), posing challenges for the application of these methods.

Recently, researchers have attempted to simulate real-world rPPG measurement scenarios through test-time adaptation (TTA) \cite{XieSFDArPPG,LiBiTTA2024,HuangFTTArPPG2024}. As shown in Fig. \ref{fig:comparsion_abc}b, TTA methods perform online unsupervised updates of the rPPG model at test-time to eliminate the distribution difference between the source and target domains. When the target domain remains static, TTA methods can ensure stable optimization for rPPG measurement as the model is continually updated. However, the assumption that the target domain remains static is often violated in real-world scenarios. In application scenarios like remote health monitoring and human-computer interaction, the dynamic environments (e.g., lighting, device aging, and user behavior) can lead to various characteristics in non-physiological information. These continual domain shifts typically cause cumulative erroneous updates, which is manifested as catastrophic forgetting. Additionally, since each application scenario contains different non-physiological information, the video distributions across different domains are diverse. It may lead to conflicting optimizations among dynamic domains, which is manifested as the over-adapted model losing its ability to adapt to future domains.

To address the limitations of catastrophic forgetting and over-adaptation, this paper proposes a stable continual test-time adaptation (CTTA) framework for rPPG measurement, whose conceptual procedures are shown in Fig. \ref{fig:teaser}. We design this framework to \textbf{R}emember the past, \textbf{A}dapt to the current domain, and \textbf{P}reempt the future, which we refer to as \textbf{PhysRAP}. To the best of our knowledge, PhysRAP is the first approach to explore continual test-time adaptation for rPPG measurement.

First, to address catastrophic forgetting, we propose a physiology-related parameters freezing strategy. Unlike updating all model parameters for adaptation, this strategy isolates physiology-related and domain-related parameters by assessing the model's uncertainty due to dynamic domain shifts. Specifically, we calculate the uncertainty score of each model parameter with respect to the current domain and identify those parameters that are insensitive to domain shifts as the physiology-related parameter set. Furthermore, considering the correlations between parameters, we expand this set to include other parameters that are highly associated with these physiology-related parameters, thereby further protecting the physiology-related knowledge. Since these parameters are considered to contain the key knowledge for extracting physiological information from videos, we freeze them during adaptation to prevent catastrophic forgetting. Second, to address over-adaptation, we design a preemptive gradient modification strategy, which adjusts the current adaptation by pre-adapting to a potential future domain. Specifically, we apply different data augmentations to incoming video samples, treating these augmented videos as a potential future domain. Then, we attempt to adapt the rPPG model to this domain to obtain the corresponding optimization gradients. We carefully analyze the impact of the future domain on the current adaptation and design corresponding modification principles, thereby using the optimization gradients from the future domain to correct over-adaptation. Finally, in dynamic domain shifts, PhysRAP freezes the physiology-related parameter set and updates the rPPG model using the modified gradients. Leveraging the aforementioned strategies, PhysRAP is endowed with a broader range of observational capabilities, thereby providing stable and accurate rPPG measurement under dynamic domain shifts.

Our main contributions are summarized as follows: 1) We propose a stable continual test-time adaptation framework (PhysRAP) that follows the "remember-adapt-preempt" paradigm, providing stable and accurate rPPG measurement in dynamic domain shifts. 2) We propose evaluating the domain uncertainty score and then separating the model's physiology-related and domain-related parameters, thereby preventing catastrophic forgetting by freezing the physiology-related parameters. 3) We design a novel preemptive gradient modification strategy that performs pre-adaptation to a potential future domain and modifies the current adaptation accordingly, thereby preventing over-adaptation. 4) Experiments on benchmark datasets fully demonstrate the effectiveness of our method, which performs exceptionally well in dealing with continual domain shifts and achieves significant improvements.

\section{Related Work}

\begin{figure*}[htb]
\centering
\includegraphics[width=1\linewidth]{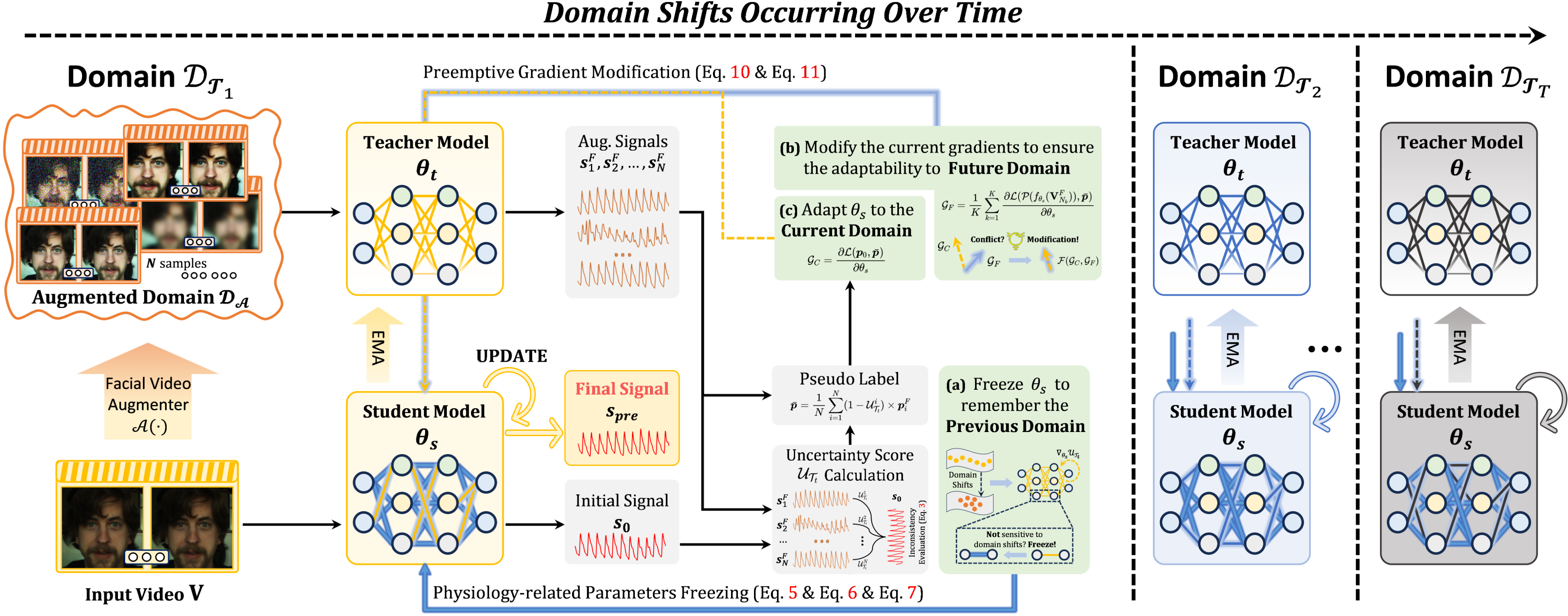}
\caption{Overview of our stable continual test-time adaptation framework for remote physiological measurement (PhysRAP), which continually adapts a base rPPG model, trained on a source domain $\mathcal{D_S}$, to $T$ target domains $\mathcal{D}_{\mathcal{T}_1},\mathcal{D}_{\mathcal{T}_2},\dots,\mathcal{D}_{\mathcal{T}_T}$.}
\label{fig:overview}
\Description{..}\vspace{-1em}
\end{figure*}

\subsection{Remote Physiological Measurement}
\label{subsec:physiological_measurement}

Remote physiological measurement aims to mine the periodic light absorption changes caused by heartbeats in facial videos. Since the early study reported in \cite{VerkruysseGreen2008}, numerous rPPG methods have been developed. Traditional signal processing methods are typically built on color space transformation \cite{WangPOS2017,DeCHROM2013} and signal decomposition \cite{PohICA2010,LewandowskaPCA2011}. However, due to the strong assumptions they rely on, these methods may not perform well in complex scenarios. With the development of deep learning (DL), DL-based models \cite{YuPhysNet2019,DasBVP2021,NiuRhythm2020,NiuCVD2020,shao2023tranphys,YuPhysFormer2022,YuPhysformer++2023,qian2025physdiff} have become increasingly prominent in rPPG measurement. Among these methods, Transformer-based approaches \cite{YuPhysFormer2022,qian2024dual,YuPhysformer++2023,LiurPPGMAE2023,shao2023tranphys,Dual-TL2024}, which can extract global information from remote facial videos, have gradually become dominant. Despite these methods have made significant progress, most of them are based on the unrealistic assumption that the source and target domains are identical. 

\subsection{Test-time Adaptation (TTA)}
\label{subsec:tta}

Test-time adaptation (TTA) aims to address domain shifts between training and test videos, which belongs to the source-free domain adaptation paradigm \cite{HuTTA12021,LiBiTTA2024,XieSFDArPPG,NiuETTA2022,HuangFTTArPPG2024,guo2024benchmarking}. Unlike standard unsupervised domain adaptation, which requires training data access, TTA methods typically use teacher-student networks to generate pseudo-labels for unsupervised updating. TTA models often focus on improving pseudo-label quality. For instance, AdaContrast \cite{ChenAdaContrast2022} used weak and strong augmentations for contrastive learning to refine pseudo-labels. SFDA-rPPG \cite{XieSFDArPPG} employed various spatiotemporal augmentations to enhance pseudo-label quality. However, they require the test data to maintain an unchanging distribution, which is usually not guaranteed in real-world scenarios.

\subsection{Continual Test-time Adaptation (CTTA)}
\label{subsec:cotta}

Continual test-time adaptation (CTTA) targets a realistic scenario where the target domain distribution shifts over time during testing. CoTTA \cite{WangCoTTA2022} first introduced this scenario and established a teacher-student network baseline. Most subsequent CTTA methods focused on mitigating catastrophic forgetting. Specifically, PETAL \cite{BrahmaPETAL2023}, HoCoTTA \cite{CuiHoCoTTA2024}, and SRTTA \cite{DengSRTTA2023} isolated the domain-invariant parameters using the Fisher information matrix to prevent error accumulation during continual adaptation. DA-TTA \cite{WangDaTTA2024} and RoTTA \cite{YuanRoTTA2023} proposed updating only batch normalization parameters to avoid model drift. Our PhysRAP further enhances the model's adaptability to future domains, achieving stable rPPG measurement by comprehensively considering the past, present, and future.

\section{Methodology}
\label{sec:method}

\subsection{Problem Definition}
\label{subsec:preliminary}

Given facial videos $(\mathbf{V}_1,\mathbf{V}_2,\dots,\mathbf{V}_n)\in\mathcal{D_S}$ and the corresponding ground-truth PPG signals $(\boldsymbol{s}_1,\boldsymbol{s}_2,\dots,\boldsymbol{s}_n)\in\mathcal{Y_S}$ collected in a specific scenario, existing methods typically train the rPPG model $f_{\theta}:\mathbf{V}\to\boldsymbol{s}$ on the source domain $\mathcal{D_S}$ and deploy it to the target domain $\mathcal{D_T}$, under the assumption of consistent data distribution (i.e., $\mathcal{D_S}=\mathcal{D_T}$). Recent rPPG researches \cite{LeeMeta2020,HuangFTTArPPG2024,XieSFDArPPG,LiBiTTA2024} challenge this assumption in real-world scenarios, where $\mathcal{D_S}\neq\mathcal{D_T}$, proposing the \textit{deploying-and-adapting} strategy. In this setting, the rPPG model $f_\theta$ updates itself based on incoming facial videos, without using any source video from $\mathcal{D_S}$. However, these works are still limited by the ideal assumption that the target domain remains static after deployment, i.e., $\mathcal{D}_{\mathcal{T}_1}=\mathcal{D}_{\mathcal{T}_2}=\dots=\mathcal{D}_{\mathcal{T}_T}$.

Motivated by the dynamic individual behavior patterns and video collection environments, our work introduces a more realistic scenario for deploying rPPG models. In this scenario, the target domain differs from the source domain (i.e., $\mathcal{D_S}\neq\mathcal{D}_{\mathcal{T}_{1:T}}$) and its data distribution changes continually, i.e., $\mathcal{D}_{\mathcal{T}_1}\neq\mathcal{D}_{\mathcal{T}_2}\neq\dots\neq\mathcal{D}_{\mathcal{T}_T}, T>1$.

\subsection{Overall Framework}
\label{subsec:overall}

As shown in Fig. \ref{fig:overview}, the framework of PhysRAP starts with a pre-trained rPPG measurement teacher-student model, $\theta_t$ and $\theta_s$, both of which have the same network structure. Given the testing video sample $\mathbf{V}$ from a novel domain $\mathcal{D}_{\mathcal{T}_t}$, PhysRAP aims to adapt the student model $\theta_s$ to the distribution of $\mathbf{V}$ in an unsupervised manner, thereby updating $\theta_s$ and obtaining the rPPG signal $\boldsymbol{s}_{pre}$. 

To ensure continual and stable rPPG measurement, PhysRAP is required to address the inevitable issues of catastrophic forgetting and over-adaptation in dynamic environments. Therefore, we design the procedures of PhysRAP from three aspects: remembering the physiology-related knowledge (embodied in Fig. \ref{fig:overview}a), preserving the ability to adapt to future domains (embodied in Fig. \ref{fig:overview}b), and adapting to the current domain (embodied in Fig. \ref{fig:overview}c).

Specifically, PhysRAP initially conducts \textbf{Domain Uncertainty Score Calculation} of $\theta_s$ in the current domain using facial video augmenter $\mathcal{A}(\cdot)$ and $\theta_t$. Subsequently, the uncertainty score is utilized for \textbf{Physiology-related Parameters Identification}, which enables the separation of physiology-related and domain-related parameters. These physiology-related parameters are considered essential for retaining the capability for rPPG measurement. \textit{Accurately identifying and freezing these parameters during adaptation is the key insight for PhysRAP to prevent catastrophic forgetting.} Next, PhysRAP simulates a potential future domain and performs \textbf{Future Domain Pre-adaptation}, using gradients from the future domain to modify the current adaptation. \textit{Preemptively adapting to potential future domains and modifying the current adaptation accordingly is the key insight for PhysRAP to prevent over-adaptation.} Finally, PhysRAP executes \textbf{Stable Test-time Adaptation} by updating the model using modified gradients while freezing the physiology-related parameters. Overall, PhysRAP integrates considerations for previous, current, and future domains during adaptation. This approach avoids catastrophic forgetting and over-adaptation while ensuring stable and accurate rPPG measurements.

\subsection{Physiology-related Parameters Freezing}
\label{subsec:physiology-related_param}
Usually, the effectiveness of rPPG models hinges on two key factors: (i) identifying rPPG signal patterns in facial areas (i.e., physiology-related knowledge) and (ii) minimizing the interference from non-physiological information (i.e., domain-related knowledge). In dynamic environments, rPPG models are prone to catastrophic forgetting due to the accumulation of erroneous updates that modify the physiology-related knowledge. Therefore, to mitigate the accumulation of errors and catastrophic forgetting, it is necessary to identify different knowledge and utilize them separately.

To this end, we propose freezing the parameters that retain physiology-related knowledge during adaptation. The Fisher Information Matrix (FIM) has been proven to effectively measure the sensitivity of model parameters to new domains based on their domain uncertainty \cite{SpallFIM2003,BrahmaPETAL2023,DengSRTTA2023}. We denote the sensitivity of model parameters to domain shift as the domain uncertainty score and identify those parameters that are insensitive to domain shifts as physiology-related parameters. Therefore, this process essentially consists of two parts: (i) calculating the domain uncertainty score $\mathcal{U}_{\mathcal{T}_t}$ and (ii) identifying the physiology-related parameter set $\mathcal{I_P}$.

\subsubsection{Domain Uncertainty Score Calculation}
\label{subsec:stable_score}

Previous works \cite{AnttiMeanTeacher2017,MarRMT2023} have demonstrated that the mean teacher predictions can provide stable pseudo-labels in dynamic environments. Based on this insight, we evaluate the consistency of the rPPG signals from augmented videos. The greater the deviation of rPPG signals from augmented videos compared to the original video, the higher the model's domain uncertainty, and vice versa. This relationship helps in identifying the physiology-related and domain-related parameters. Therefore, the key to calculating the domain uncertainty score $\mathcal{U}_{\mathcal{T}_t}\in\mathbb{R}$ is to assess the consistency of the teacher model's predictions.

Concretely, given the input facial video $\mathbf{V}\in\mathbb{R}^{3\times D\times H\times W}$, to calculate the domain uncertainty, we first apply perturbations to the video $\mathbf{V}$ in the current domain $\mathcal{D}_{\mathcal{T}_t}$ using a facial video augmenter $\mathcal{A}(\cdot)$, which generates $N$ augmented video samples:
\begin{equation}
    \label{eq:get_aug_video}
    \mathbf{V}_1^F,\mathbf{V}_2^F,\dots,\mathbf{V}_N^F=\mathcal{A}(\mathbf{V}),
\end{equation}
where $\mathcal{A}(\cdot)$ generates $N$ videos by randomly selected augmentation methods, and $D$, $H$, $W$ refer to the length, height, and width of the input video, respectively. After that, the reference signal $\boldsymbol{s}_{0}\in\mathbb{R}^{D}$ and the augmented signals $\boldsymbol{s}^F_1,\boldsymbol{s}^F_2,\dots,\boldsymbol{s}^F_N$   could be obtained by:
\begin{equation}
    \label{eq:get_aug_signal}
    \begin{split}
        \boldsymbol{s}_{0}=f_{\theta_s}(\mathbf{V}),\ \boldsymbol{s}_i^F=f_{\theta_t}(\mathbf{V}_i^F).
    \end{split}
\end{equation}

As we just discussed, the inconsistency between these augmented signals and the reference signal can be used to measure the model's uncertainty score $\mathcal{U}_{\mathcal{T}_t}$ in the current domain $\mathcal{D}_{{\mathcal{T}_t}}$: 
\begin{equation}
    \label{eq:get_us}
    \begin{split}
        &\mathcal{U}_{\mathcal{T}_t}=\frac{1}{N}\sum_{i=1}^N\mathcal{U}_{\mathcal{T}_t}^i=\frac1{N}\sum_{i=1}^N\\&\frac{1}{2}\left(\underbrace{1-\exp\left (-\left |\frac{\boldsymbol{s}_{0}}{\|\boldsymbol{s}_{0}\|}-\frac{\boldsymbol{s}_i^F}{\|\boldsymbol{s}_i^F\|} \right |\right )}_{temporal}+\underbrace{1-\exp\left (-\left |\frac{\boldsymbol{p}_{0}}{\|\boldsymbol{p}_{0}\|}-\frac{\boldsymbol{p}_i^F}{\|\boldsymbol{p}_i^F\|} \right |\right )}_{frequency}\right),
    \end{split}
\end{equation}
where $\boldsymbol{p}_0=\mathcal{P}(\boldsymbol{s}_0),\boldsymbol{p}_i^F=\mathcal{P}(\boldsymbol{s}_i^F)$ and $\mathcal{P}(\cdot)$ denotes the calculation of power spectral density. Generally, the domain uncertainty $\mathcal{U}_{\mathcal{T}_t}\in[0,1]$ reflects the uncertainty of $\theta_s$ with respect to domain $\mathcal{D}_{{\mathcal{T}_t}}$ by comprehensively evaluating the inconsistency of the teacher's predictions in both the temporal and frequency domains.

\subsubsection{Physiology-related Parameters Identification}
\label{subsec:stable_score}

\begin{figure}[t]
\centering
\includegraphics[width=1\columnwidth]{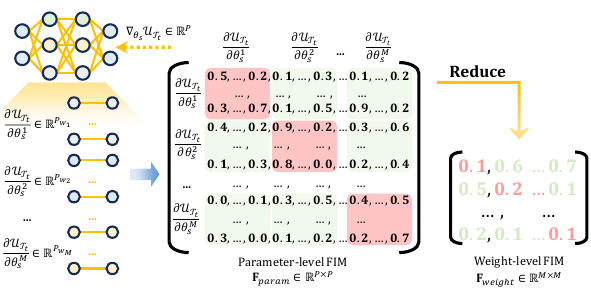}
\caption{Visualization of the calculation of weight-level fisher information matrix $\mathbf{F}_{weight}$.}
\Description{..}\vspace{-1em}
\label{fig:fim}
\end{figure}

Recent pioneer CTTA methods \cite{WangCoTTA2022,BrahmaPETAL2023,CuiHoCoTTA2024} have already successfully estimated the sensitivity of parameters by computing the parameter-level Fisher information matrix $\mathbf{F}_{param}$ from the domain uncertainty score $\mathcal{U}_{{\mathcal{T}_t}}$:
\begin{equation}
    \label{eq:get_f_param}
    \mathbf{F}_{param}=\left(\frac{\partial \mathcal{U}_{{\mathcal{T}_t}}}{\partial \theta_s}\right)\left( \frac{\partial \mathcal{U}_{{\mathcal{T}_t}}}{\partial \theta_s}\right)^\top,
\end{equation}
where $\mathbf{F}_{param}\in\mathbb{R}^{P\times P}$, $P$ is the number of parameters in $\theta_s$. The diagonal elements of $\mathbf{F}_{param}$ denotes the sensitivity of parameters, while the non-diagonal elements represent the correlation of them.

However, due to the large number of parameters in $\theta_s$, it is impractical to calculate all elements in $\mathbf{F}_{param}$. Existing methods usually consider the diagonal elements (i.e., sensitivity) but neglect the non-diagonal elements (i.e., correlation). Therefore, for lower calculation and higher explainability, we propose to reduce $\mathbf{F}_{param}$ to the weight-level FIM $\mathbf{F}_{weight}\in\mathbb{R}^{M\times M}$, where $M$ is the number of weights in $\theta_s$. The weight is defined as the relevant parameters within a functional module, which collectively achieve a particular computational function. For example, the parameters of a convolutional kernel, the bias vector, and so on.

As shown in Fig. \ref{fig:fim}, the number of weights $M$ is much smaller than the number of parameters \( P \) (\( M \ll P \)), which makes it possible to compute all the elements in $\mathbf{F}_{weight}$. This allows us to comprehensively consider the sensitivity and correlation of the parameters, thereby obtaining more accurate physiology-related parameter set $\mathcal{I_P}$. Formally, the weight-level FIM could be obtained by:
\begin{equation}
\label{eq:get_f}
\begin{split}
	\mathbf{F}^{i,j}_{weight}=\begin{cases}
    \frac{1}{P_{w_i}P_{w_j}}\sum_{l=1}^{P_{w_i}}\sum_{k=1}^{P_{w_j}}\left (\frac{\partial  \mathcal{U}_{\mathcal{T}_t}}{\partial \theta_s^i}\right )\left (\frac{\partial  \mathcal{U}_{\mathcal{T}_t}}{\partial \theta_s^j}\right )^\top_{kl}\ \mathrm{if}\ i\neq j\\
	\frac{1}{P_{w_i}}\sum_{k=1}^{P_{w_i}}\left (\frac{\partial  \mathcal{U}_{\mathcal{T}_t}}{\partial \theta_s^i}\right )^2_k\ \mathrm{otherwise},	
	\end{cases}
\end{split}
\end{equation}
where $P_{w_i}$ denotes the number of parameters in the $i$-th weight $\theta_s^i$ and $P=\sum_{i=1}^MP_{w_i}$. Note that $\mathbf{F}^{i,j}_{weight}$ denotes the correlation between $\theta_s^i$ and $\theta_s^j$, while $\mathbf{F}_{weight}^{i,i}$ denotes the sensitivity of $\theta_s^i$. 

Subsequently, we obtain the physiology-related parameter set $\mathcal{I_P}$ in two steps. We firstly initialize $\mathcal{I_P}$ with the $r_1\%$ of weights that are the least sensitive for domain shifts brought by $\mathcal{D}_{\mathcal{T}_t}$:
\begin{equation}
\label{eq:ip_init}
    \mathcal{I_P}=\left\{\theta_s^i|\ \mathbf{F}_{weight}^{ii}< Top_{(1-r_1\%)}\left(\mathrm{Diag(\mathbf{F}_{weight})}\right)\right\}.
\end{equation}
Afterward, we expand the elements of $\mathcal{I_P}$ to include the weights whose correlation with each element in $\mathcal{I_P}$ is in the top $r_2\%$:
\begin{equation}
\label{eq:ip_expand}
    \mathcal{I_P}=\bigcup_{\theta_s^i\in\mathcal{I_P}}\left\{\theta_s^j|\ \mathbf{F}^{ij}_{weight}>Top_{(r_2\%)}\mathbf{F}_{weight}^i\right\}\cup\mathcal{I_P},
\end{equation}
where $\mathbf{F}_{weight}^i$ denotes the $i$-th line of $\mathbf{F}_{weight}$. In summary, the physiology-related parameter set $\mathcal{I_P}$ not only includes parameters that are insensitive to domain shifts but also those that are highly correlated with these physiology-related parameters. This strategy protects the model's ability to extract physiological information during the adaptation.
\subsection{Future Domain Pre-adaptation}
\label{subsec:over-adaptation}
\begin{figure}[t]
\centering
\includegraphics[width=1\columnwidth]{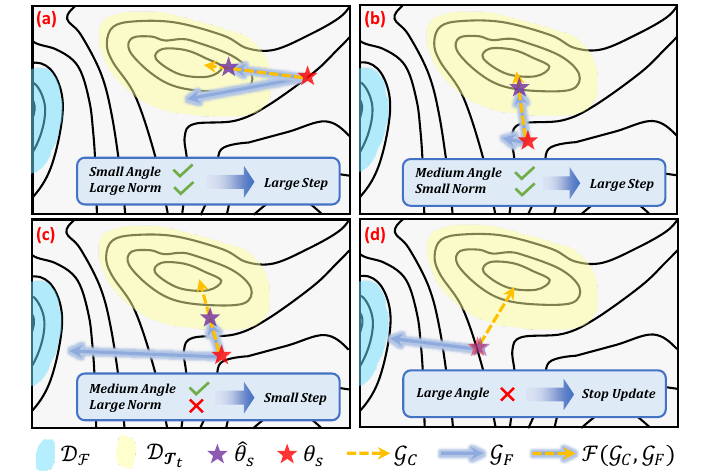}
\caption{Visualization of four examples for the preemptive gradient modification strategy.}
\label{fig:gradient}\vspace{-1.5em}
\Description{..}
\end{figure}

Previous works \cite{WangNTNIPS2021,WangNTTGRS2022,ZhouNTAAAI2023,JiangNTNIPS2023} in continual learning have found that models over-fitted to the source domain are difficult to transfer to new domains. This issue is summarized as negative transfer, which may arise from conflicts in optimal weight configurations caused by dynamic data distributions. Therefore, in CTTA, we speculate that if $\theta_s$ over-adapts to the current domain, it may similarly undermine the adaptability to future domains. Based on this insight, we suggest that $\theta_s$ should update its parameters after preemptively taking into account its adaptability to future domains.

To achieve the above goals, we design a preemptive gradient modification strategy. It proactively simulates a potential future domain $\mathcal{D_F}=\{\mathbf{V}_{N_k}^F\}_{k=1}^K,N_k\in[1,N]$, which is a subset of the augmented domain $\mathcal{D_A}$. Afterward, we perform the pre-adaptation to this potential future domain to obtain the future optimization gradient $\mathcal{G}_F$, which could be used to modify the current optimization gradient $\mathcal{G}_C$, thereby ensuring adaptability to future domains. Specifically, the future domain pre-adaptation process consists of three steps: (i) pseudo-label computation, (ii) optimization gradient calculation, and (iii) gradient correction.

First, considering the domain uncertainty score of $\mathbf{V}_i^F$ reflects the confidence of the corresponding PSD signal $\boldsymbol{p}_i^F$, we aggregate the PSD signals from augmented videos based on their uncertainty scores: 
\begin{equation}
    \label{eq:get_pesudo}
    \bar{\boldsymbol{p}}=\frac{1}{N}\sum_{i=1}^N (1-\mathcal{U}_{\mathcal{T}_t}^i)\times\boldsymbol{p}_i^F.
\end{equation}

Second, we use backpropagation to separately compute the optimization gradients for $\theta_s$ to adapt to the current domain $\mathcal{D}_{\mathcal{T}_t}$ and the future domain $\mathcal{D_F}$:
\begin{equation}
    \label{eq:get_gradient}
    \mathcal{G}_C=\frac{\partial \mathcal{L}(\boldsymbol{p}_0,\bar{\boldsymbol{p}})}{\partial \theta_s},\ \ \mathcal{G}_F=\frac{1}{K}\sum_{k=1}^K\frac{\partial \mathcal{L}(\mathcal{P}(f_{\theta_s}(\mathbf{V}_{N_k}^F)),\bar{\boldsymbol{p}})}{\partial \theta_s},
\end{equation}
where $\mathcal{L}(\cdot,\cdot)$ denotes the cross-entropy loss, $\mathcal{P}(\cdot)$ denotes the PSD calculation, and $\mathcal{G}_C,\mathcal{G}_F\in\mathbb{R}^{P}$ are the gradients for each parameters of the student model $\theta_s$.

Finally, we modify $\mathcal{G}_C$ according to the preemptive gradient modification strategy, which takes into account both the norm and direction of $\mathcal{G}_F$ and performs weight-level gradient modification for $\theta_s$. Specifically, for the gradient $\mathcal{G}_C^i,\mathcal{G}_F^i\in\mathbb{R}^{P_{w_i}}$ corresponding to the $i$-th weight $\theta_s^i$, we perform the modification following three principles: (i) When the directions of $\mathcal{G}_C^i$ and $\mathcal{G}_F^i$ are considered to be non-conflicting, a large step size update can be performed, as illustrated in Fig. \ref{fig:gradient}a; (ii) When there is some conflict between the directions of  $\mathcal{G}_C^i$ and $\mathcal{G}_F^i$, the step size of the update decreases as the magnitude of $\mathcal{G}_F^i$ increases, as illustrated in Fig. \ref{fig:gradient}b and Fig. \ref{fig:gradient}c; (iii) When there is severe conflict between the directions of $\mathcal{G}_C^i$ and $\mathcal{G}_F^i$, adaptation according to $\mathcal{G}_C^i$ should be stopped, as illustrated in Fig. \ref{fig:gradient}d. Note that we use the degree of the angle between directions to distinguish their conflict level. We set the boundaries for small, medium, and large angles at 0, \(\pi/4\), \(\pi/2\), and $\pi$, respectively.

Based on the above design, we modify the gradients $\mathcal{G}_C^i,\mathcal{G}_F^i$ according to the following formula:
\begin{equation}
    \label{eq:modify_gc}
    \mathcal{F}(\mathcal{G}_C^i,\mathcal{G}_F^i)=\begin{cases}
        0\ \ \mathrm{if}\ \cos\langle\mathcal G_F^i,\mathcal G_C^i \rangle <0\\
        w_i\times\mathcal{G}_C^i\ \ \mathrm{otherwise,}
    \end{cases}
\end{equation}
where $w_i\in(0,1)$ denotes the correction coefficient for the current gradient, which could be calculated by:
\begin{equation}
    \label{eq:get_w_i}
    \begin{split}
        w_i=&\frac{1}{1+\exp\left(-\frac{\|\mathcal G_F^i\|}{\|\mathcal G_C^i\|}\left(\cos\langle\mathcal G_F^i,\mathcal G_C^i \rangle-\frac{\sqrt 2}{2}\right) \right)}.
    \end{split}
\end{equation}

\subsection{Stable Test-time Adaptation}
\label{subsec:preliminary}

In the preceding steps, we have already identified the physiology-related parameter set $\mathcal{I_P}$ to prevent catastrophic forgetting and obtained the modified gradients $\mathcal{F}(\mathcal{G}_C^i,\mathcal{G}_F^i)$ to prevent over-adaptation. We employ these designs to ensure that PhysRAP can perform rPPG measurements in a continual and stable manner. The student model $\theta_s$ is updated with the following formula:
\begin{equation}
    \label{eq:update_s}
    \hat\theta_s^{i}\leftarrow
    \begin{cases}
        \theta_s^i-\eta\mathcal{F}(\mathcal{G}_C^i,\mathcal{G}_F^i),\ \mathrm{if}\ \theta_s^i\notin\mathcal{I_P}\\
        \theta_s^i,\ \mathrm{otherwise},\\
    \end{cases}
\end{equation}
where $\eta=1e-4$ denotes the learning rate and $\hat\theta_s^{i}$ denotes the updated parameters of $i$-th weight. Note that only the parameters not belonging to the physiology-related parameter set $\mathcal{I_P}$ are updated. 

Subsequently, the corresponding parameters of teacher model $\theta_t$ are updated by the widely-used exponential moving average (EMA) to ensure maximal model plasticity:
\begin{equation}
    \label{eq:update_t}
    \hat\theta_t\leftarrow \alpha\theta_t+(1-\alpha)\hat\theta_s,
\end{equation}
where $\alpha=0.99$ denotes the momentum factor. Afterward, when the model receives the next video sample $\mathbf{V}'$, it will repeat all the aforementioned procedures, with parameters initialized as $\theta_s\leftarrow \hat{\theta}_s,\theta_t\leftarrow\hat{\theta}_t$. Generally, we summarize our proposed stable continual test-time adaptation framework PhysRAP in Algorithm \ref{alg:cap}.

\begin{algorithm}
  \caption{Stable CTTA Framework for rPPG Measurement}\label{alg:cap}
  \begin{algorithmic}[1]
    \Require facial video augmenter $\mathcal{A}(\cdot)$, student model $\theta_s$, teacher model $\theta_t$, learning rate $\eta$, momentum factor $\alpha$, parameters frozen ratios $r_1\%,r_2\%$, videos from current domain $\mathcal{D}_{\mathcal{T}_t}$;
    \Ensure final rPPG signal $\boldsymbol{s}_{pre}$;
    \For{each $\mathbf{V}\in\mathcal{D}_{\mathcal{T}_t}$}
        \State \Comment{Domain Uncertainty Score Calculation}
        \State Generate augmented videos $\{\mathbf{V}_i^F\}_{i=1}^N=\mathcal{A}(\mathbf{V})$;
        \State Compute initial signals $\boldsymbol{s}_{0}=f_{\theta_s}(\mathbf{V}),\ \boldsymbol{s}_i^F=f_{\theta_t}(\mathbf{V}_i^F)$;
        \State Compute the uncertainty score $\mathcal{U}_{\mathcal{T}_t}$ with Eq. \ref{eq:get_us};
        \State \Comment{Physiology-related Parameters Identification}
        \State Compute weight-level FIM $\mathbf{F}$ with Eq. \ref{eq:get_f};
        \State Initialize and expand $\mathcal{I_P}$ with Eq. \ref{eq:ip_init}, \ref{eq:ip_expand};
        \State \Comment{Future Domain Pre-adaptation}
        \State Compute the pseudo-label $\bar{\boldsymbol{p}}=\frac{1}{N}\sum_{i=1}^N (1-\mathcal{U}_{\mathcal{T}_t}^i)\times\boldsymbol{p}_i^F$;
        \State Compute gradients $\mathcal{G}_C, \mathcal{G}_F$ with Eq. \ref{eq:get_gradient};
        \State Modify $\mathcal{G}_C$ to $\mathcal{F}(\mathcal{G}_C,\mathcal{G}_F)$ with Eq. \ref{eq:modify_gc}, \ref{eq:get_w_i};
        \State \Comment{Stable Test-time Adaptation}
        \State Update student model $\theta_s$ to $\hat{\theta}_s$ with Eq. \ref{eq:update_s};
        \State Update teacher model $\theta_t$ to $\hat{\theta}_t$ with Eq. \ref{eq:update_t};
        \State Make final prediction $\boldsymbol{s}_{pre}\gets f_{\hat{\theta}_s}(\mathbf{V})$;
        \State $\theta_s\gets\hat{\theta}_s,\ \theta_t\gets\hat{\theta}_t$;
    \EndFor
  \end{algorithmic}
\end{algorithm}

\section{Experimental Results}
\label{sec:experimets}

\begin{table*}[t]
\centering
\caption{HR estimation results under CTTA protocol. The symbols $\triangle$, $\star$, and $\ddagger$ denote the traditional, deep learning-based, and TTA methods (based on ResNet3D-18 \cite{HaraResNet3D2018}), respectively. The symbol $\downarrow$ indicates lower is better, and $\uparrow$ indicates higher is better. Best results are marked in \textbf{bold} and second best in \underline{underline}. The metrics M and R are short for the MAE and RMSE.}
\setlength\tabcolsep{2pt}
\begin{tabular}{c|ccc|ccc|ccc|ccc|ccc|ccc|ccc}
\toprule
Time & \multicolumn{18}{l|}{$\ t\xrightarrow{\hspace*{12cm}}$}& \\ 
\midrule
\multirow{2}{*}{Method} & \multicolumn{3}{c}{UBFC-rPPG} & \multicolumn{3}{c}{UBFC-rPPG$^+$} & \multicolumn{3}{c}{PURE} &\multicolumn{3}{c}{PURE$^+$} & \multicolumn{3}{c}{BUAA-MIHR} & \multicolumn{3}{c}{BUAA-MIHR$^+$} & \multicolumn{3}{c}{MEAN}\\
\cmidrule(lr){2-4} \cmidrule(lr){5-7} \cmidrule(lr){8-10} \cmidrule(lr){11-13} \cmidrule(lr){14-16} \cmidrule(lr){17-19} \cmidrule(lr){20-22}
& M$\downarrow$ & R$\downarrow$ & $r\uparrow$& M$\downarrow$ & R$\downarrow$ & $r\uparrow$& M$\downarrow$ & R$\downarrow$ & $r\uparrow$& M$\downarrow$ & R$\downarrow$ & $r\uparrow$& M$\downarrow$ & R$\downarrow$ & $r\uparrow$& M$\downarrow$ & R$\downarrow$ & $r\uparrow$& M$\downarrow$ & R$\downarrow$ & $r\uparrow$\\
\midrule
GREEN$^\triangle$\cite{VerkruysseGreen2008}&50.2&52.4&0.04&50.2&52.4&0.20&24.4&33.1&0.10&24.2&33.1&0.24&37.0&38.3&0.03&37.0&38.3&0.05&37.2&41.3&0.11\\
ICA$^\triangle$\cite{PohICA2010}&14.8&18.2&0.72&15.4&19.0&0.66&9.30&14.6&0.86&9.33&15.1&0.85&7.99&9.49&0.81&8.47&10.2&0.78&10.9&14.4&0.78\\
POS$^\triangle$\cite{WangPOS2017}&9.33&12.5&0.73&7.64&10.2&0.84&9.85&13.4&0.89&8.34&12.3&0.90&4.28&5.63&0.83&5.49&6.97&0.72&7.49&10.2&0.82\\
\midrule
PhysNet$^\star$\cite{YuPhysNet2019}&13.2&20.4&0.22&13.0&20.7&0.25&9.01&19.7&0.54&8.30&19.0&0.59&3.62&5.91&0.89&3.69&5.58&0.91&8.48&15.2&0.57\\
PhysMamba$^\star$\cite{LuoPhysMamba2024}&19.2&26.7&0.30&11.8&19.9&0.29&6.84&18.6&0.65&6.90&18.7&0.65&4.06&6.22&0.89&3.77&6.31&0.87&8.76&16.1&0.61\\
PhysFormer$^\star$\cite{YuPhysFormer2022}&1.78&2.97&\underline{0.98}&2.66&6.43&0.91&7.99&16.7&0.69&7.88&16.0&0.72&3.45&5.18&0.92&3.62&5.54&0.89&4.56&8.82&0.85\\
RhythmMamba$^\star$\cite{zou2024rhythmmamba}&2.65&2.53&\textbf{0.99}&3.12&6.22&0.92&1.99&4.54&0.98&2.06&3.21&0.99&4.65&7.05&0.82&4.18&5.47&0.90&3.10&4.83&0.93\\
\midrule
CoTTA$^{\ddagger\star}$\cite{WangCoTTA2022}&\underline{1.43}&\underline{2.48}&\textbf{0.99}&1.46&3.89&\underline{0.97}&\textbf{0.58}&\textbf{1.41}&\textbf{1.00}&3.55&13.5&0.82&10.8&17.0&0.06&23.9&27.5&0.16&6.96&11.0&0.67\\
DA-TTA$^{\ddagger\star}$\cite{WangDaTTA2024}&1.51&2.59&\textbf{0.99}&1.89&5.21&0.94&4.18&11.5&0.87&3.07&9.07&0.93&2.78&3.95&\underline{0.96}&4.18&6.89&0.89&3.23&7.90&0.91\\
RoTTA$^{\ddagger\star}$\cite{YuanRoTTA2023}&1.80&3.15&\underline{0.98}&3.56&8.84&0.84&3.51&9.38&0.92&9.80&19.1&0.26&2.73&3.96&\underline{0.96}&3.57&5.29&0.91&4.16&8.29&0.86\\
PETAL$^{\ddagger\star}$\cite{BrahmaPETAL2023}&1.69&2.99&\underline{0.98}&2.82&7.98&0.96&2.02&5.54&0.97&4.23&12.0&0.86&\underline{2.64}&\underline{3.81}&\underline{0.96}&\underline{2.82}&\underline{4.18}&\underline{0.95}&2.70&6.09&0.93\\
\midrule
Baseline$^{\ddagger\star}$\cite{HaraResNet3D2018}&1.78&2.97&\underline{0.98}&2.66&6.43&0.91&7.99&16.7&0.69&7.88&16.1&0.72&3.42&4.79&0.94&3.62&5.54&0.89&4.55&8.76&0.86\\
\textbf{Ours} w/o $\mathcal{I_P},\mathcal{F}^{\ddagger\star}$&1.46&2.58&\textbf{0.99}&1.54&2.65&\textbf{0.99}&1.49&4.44&0.98&\underline{0.44}&\underline{0.94}&\textbf{1.00}&7.27&13.0&0.37&7.87&12.0&0.51&3.35&5.93&0.81\\
\textbf{Ours} w/o $\mathcal{F}^{\ddagger\star}$&1.44&\underline{2.48}&\textbf{0.99}&\underline{1.39}&\underline{2.19}&\textbf{0.99}&1.19&\underline{3.67}&\underline{0.99}&1.24&3.58&\underline{0.99}&3.27&4.93&0.93&3.62&5.51&0.91&\underline{2.02}&3.72&\underline{0.97}\\
\textbf{Ours} w/o $\mathcal{I_P}^{\ddagger\star}$&1.60&2.70&\textbf{0.99}&1.41&2.27&\textbf{0.99}&1.17&3.77&\underline{0.99}&0.52&1.55&\textbf{1.00}&3.15&4.92&0.93&4.39&6.92&0.85&2.04&\underline{3.69}&0.96\\
\textbf{PhysRAP(ours)}$^{\ddagger\star}$ & {\textbf{0.81}} & {\textbf{1.84}} & {\textbf{0.99}}& {\textbf{0.85}} & {\textbf{2.13}} & {\textbf{0.99}}& {\underline{1.10}} & {3.78} & \underline{0.99} & \textbf{{0.31}} & \textbf{{0.75}} & \textbf{{1.00}}& \textbf{{2.46}} & \textbf{{3.33}} & \textbf{{0.98}}& \textbf{{2.48}} & \textbf{{3.65}}& \textbf{0.96} & \textbf{{1.33}} & \textbf{{2.58}} &\textbf{0.98}\\

\bottomrule
\end{tabular}
\label{table:dynamic_test}
\end{table*}

\subsection{Datasets and Performance Metrics}
\label{subsec:Datasets}

To demonstrate the persistent adaptability of PhysRAP, we select four datasets and expand three additional datasets based on them with data augmentation algorithms. The heart rate estimation evaluation are executed on these seven datasets.

\textbf{VIPL-HR} \cite{NiuVIPL2018} is a challenging large-scale dataset for rPPG measurement, which contains 2,378 RGB videos of 107 subjects. \textbf{UBFC-rPPG} \cite{BobbiaUBFC2019} includes 42 RGB videos recorded at a frame rate of 30 fps, captured under sunlight and indoor illumination conditions. \textbf{PURE} \cite{StrickerPURE2014} contains 60 RGB videos from 10 subjects, involving six different head motion tasks. \textbf{BUAA-MIHR} \cite{XiBUAA2020} is a dataset collected under various lighting conditions, and we only select the data with the luminance of 10 or higher. \textbf{UBFC-rPPG$^+$, PURE$^+$, and BUAA-MIHR$^+$} are augmented from the corresponding datasets with flipping, gamma correction, Gaussian blurring, and cropping.

\subsection{Evaluation Protocol}
\label{subsec:protocols}

The CTTA protocol aims to assess the unsupervised adaptation capability of pre-trained rPPG models in unknown dynamic domains. To ensure that the pre-trained model has sufficient rPPG measurement capability, we select the largest-scale VIPL-HR as the source domain $\mathcal{D_S}$ and continually adapt the rPPG model to the target domains $\mathcal{D}_{\mathcal{T}}$, where $\mathcal{D}_{\mathcal{T}}$ is the collection of the remaining six datasets. We calculate the video-level mean absolute error (MAE), root mean square error (RMSE), standard deviation of the error (SD), and Pearson's correlation coefficient ($r$) between the predicted HR and the ground-truth HR for each dataset $\mathcal{D}_{\mathcal{T}_t}\in \mathcal{D}_{\mathcal{T}}$. We leverage the average metric $\mathrm{MEAN}$ across all datasets to evaluate the model's continual adaptation ability:
\begin{equation}
    \label{eq:get_eval_a}
    \mathrm{MEAN}_{\mathrm{m}\in\{\mathrm{SD,MAE,RMSE},r\}}=\frac{1}{|\mathcal{D}_{\mathcal{T}}|}\sum_{\mathcal{D}_{\mathcal{T}_t}\in \mathcal{D}_{\mathcal{T}}} m.
\end{equation}

\subsection{Implementation Details}
\label{subsec:implementation}

We implement our proposed PhysRAP with PyTorch framework on one 24G RTX3090 GPU. Following \cite{YuPhysFormer2022,LuNEST2023}, we use the FAN face detector \cite{BulatFAN2017} to detect the coordinates of 81 facial landmarks in each video frame. Afterward, we crop and align the facial video frames to 128$\times$128 pixels according to the obtained landmarks. The frame rate of each video is uniformly standardized to 30 fps for efficiency. We employ ResNet3D-18 \cite{HaraResNet3D2018} as the baseline model and design a separate prediction head for the rPPG signal, which consists of one point-wise 3D convolution layer and one max-pooling layer.

During the training phase, we train the baseline model for 10 epochs using the Adam optimizer \cite{KingmaADAM2015}, with the base learning rate and weight decay set to 1e-4 and 5e-5. During the CTTA phase, the augmenter $\mathcal{A}(\cdot)$ randomly performs Gaussian noise, cropping, flipping, and temporal reversal. The number of frames $D$ and the batch size are set to 160 and 4, across all phases. Hyperparameters $N$, $K$, $r_1$, and $r_2$ are set to 10, 4, 80, and 20, respectively.

\subsection{Main Results}
\label{subsec:dynamic_testing}

\begin{table*}
    \centering
    \caption{Impact of the hyper-parameters (a) $r_1-r_2$, (b) $N-K$, and (c) $\eta-\alpha$.}\vspace{-1em}
    \begin{subtable}[t]{0.3\linewidth}
        \caption{Impact of the physiology-related parameters frozen ratio $r_1$ and $r_2$.}
        \setlength{\tabcolsep}{1.3mm}{
        \begin{tabular}{c|c||cccc}
        \toprule
        \multirow{2}{*}{$r_1$}&\multirow{2}{*}{$r_2$}&\multicolumn{4}{c}{MEAN}\\
        \cmidrule(lr){3-6} 
        &&SD$\downarrow$ & MAE$\downarrow$ & RMSE$\downarrow$ & $r\uparrow$\\
        \midrule
        70&\textbf{20}&2.54&1.46&2.93&0.97\\
        \midrule
        \multirow{3}{*}{\textbf{80}}&10&2.61&1.60&3.06&0.95\\
        &\textbf{20}&\textbf{2.19}&\textbf{1.33}&2.58&\textbf{0.98}\\
        &30&{2.48}&1.67&3.01&0.94\\
        \midrule
        90&\textbf{20}&2.83&1.95&\textbf{2.48}&0.93\\
        \bottomrule
        \end{tabular}}
        \label{table:hyper-param-a}
    \end{subtable}
    \begin{subtable}[t]{0.29\linewidth}
    \caption{Impact of the number of videos in $\mathcal{D_A}$ ($N$) and $\mathcal{D_F}$ ($K$).}
        \setlength{\tabcolsep}{1.3mm}{
        \begin{tabular}{c|c||cccc}
        \toprule
        \multirow{2}{*}{$N$}&\multirow{2}{*}{$K$}&\multicolumn{4}{c}{MEAN}\\
        \cmidrule(lr){3-6} 
        &&SD$\downarrow$ & MAE$\downarrow$ & RMSE$\downarrow$ & $r\uparrow$\\
        \midrule
        5&\textbf{4}&4.82&2.78&5.55&0.80\\
        \midrule
        \multirow{3}{*}{\textbf{10}}&2&5.08&3.06&5.94&0.76\\
        &\textbf{4}&2.19&\textbf{1.33}&2.58&\textbf{0.98}\\
        &8&3.07&1.76&3.52&0.96\\
        \midrule
        15&\textbf{4}&\textbf{2.11}&1.35&\textbf{2.23}&0.97\\
        \bottomrule
        \end{tabular}}
        \label{table:hyper-param-b}
    \end{subtable}
    \begin{subtable}[t]{0.35\linewidth}
    \caption{Impact of the learn rate $\eta$ and momentum factor $\alpha$.}
        \setlength{\tabcolsep}{1.3mm}{
        \begin{tabular}{c|c||cccc}
        \toprule
        \multirow{2}{*}{$\eta$}&\multirow{2}{*}{$\alpha$}&\multicolumn{4}{c}{MEAN}\\
        \cmidrule(lr){3-6} 
        &&SD$\downarrow$ & MAE$\downarrow$ & RMSE$\downarrow$ & $r\uparrow$\\
        \midrule
        0.00005&\textbf{0.990}&2.61&1.54&3.03&\textbf{0.98}\\
        \midrule
        \multirow{3}{*}{\textbf{0.0001}}&0.985&5.49&4.57&7.35&0.68\\
        &\textbf{0.990}&\textbf{2.19}&\textbf{1.33}&\textbf{2.58}&\textbf{0.98}\\
        &0.995&{2.32}&1.36&2.69&\textbf{0.98}\\
        \midrule
        0.0002&\textbf{0.990}&6.12&2.30&5.27&0.80\\
        \bottomrule
        \end{tabular}}
        \label{table:hyper-param-c}
    \end{subtable}
    
\end{table*}

Following the CTTA protocol, we evaluate the continual adaptation ability of PhysRAP in six datasets (i.e., UBFC-rPPG, UBFC-rPPG$^+$, PURE, PURE$^+$, BUAA-MIHR, and BUAA-MIHR$^+$). To conduct a comprehensive comparison, we compare PhysRAP with both traditional methods (i.e., GREEN \cite{VerkruysseGreen2008}, ICA \cite{PohICA2010}, and POS \cite{WangPOS2017}) and deep learning-based methods (i.e., PhysNet \cite{YuPhysNet2019}, PhysMamba \cite{LuoPhysMamba2024}, PhysFormer \cite{YuPhysFormer2022}, and RhythmMamba \cite{zou2024rhythmmamba}). Furthermore, we reproduce four CTTA methods (i.e., CoTTA \cite{WangCoTTA2022}, DA-TTA \cite{WangDaTTA2024}, RoTTA \cite{YuanRoTTA2023}, and PETAL \cite{BrahmaPETAL2023}) based on ResNet3D-18 \cite{HaraResNet3D2018}, whose implementation details are provided in the supplementary material.

As shown in Tab. \ref{table:dynamic_test}, most traditional methods and deep learning-based methods exhibit average performance across various datasets, without significant performance improvement or decline over time. Moreover, some deep learning-based methods (e.g., PhysNet and PhysMamba) perform worse than traditional methods (i.e., POS). This is because the CTTA protocol measures the domain generalization ability for them, which is not strongly correlated with their fitting ability in a single domain.

For CTTA methods, it can be observed that their overall performance is significantly better than that of deep learning-based methods, except for CoTTA \cite{WangCoTTA2022}. In fact, CoTTA demonstrates its adaptability in the early stage. It performs well in the first three domains, especially achieving the best MAE (0.58 bpm) and RMSE (1.41 bpm) in the PURE dataset. However, it experiences a significant performance degradation after the PURE$^+$, which may be caused by catastrophic forgetting due to its stochastic restoration strategy. Generally, compared to PhysRAP, these CTTA methods all show good results initially but perform sub-optimally in mid-term and late-term domain adaptation. In contrast, PhysRAP demonstrates stable and accurate rPPG measurement throughout the continual adaptation process and achieves the best mean MAE (1.33 bpm), RMSE (2.58 bpm), and $r$ (0.98), which exhibits significant improvements. We believe that the advantages stem from the proposed physiology-related parameter freezing and preemptive gradient correction strategies. These two strategies respectively mitigate the catastrophic forgetting and over-adaptation issues in the CTTA process, with their benefits particularly reflected in the stability of HR estimation under continual domain shifts.

\begin{table}[tb]
\centering
\caption{Ablation study on the identification strategy for physiology-related parameters. Diag. is short for diagonal elements.}
\vspace{-0.5em}
\begin{tabular}{l|cccc}
\toprule
\multirow{2}{*}{Identification Strategy}&\multicolumn{4}{c}{MEAN}\\
\cmidrule(lr){2-5} 
&SD$\downarrow$ & MAE$\downarrow$ & RMSE$\downarrow$ & $r\uparrow$\\
\midrule
Random Selection&{3.88}&2.61&3.97&{0.97}\\
Diag. of Weight-level FIM&3.00&1.65&3.46&0.96\\
Weight-level FIM&\textbf{2.19}&\textbf{1.33}&\textbf{2.58}&\textbf{0.98}\\
\bottomrule
\end{tabular}\vspace{-0.5em}
\label{table:alb-physiology-related}
\end{table}

\subsection{Ablation Studies} 


In this section, we carry out ablation studies on the hyperparameters and core components within PhysRAP. All experiments in this section follow the CTTA protocol and report the $\mathrm{MEAN}$ metric.

\subsubsection{Impact of Physiology-related Parameters Freezing} 



As discussed in Sec. \ref{subsec:physiology-related_param}, PhysRAP freezes the physiology-related parameters before adaptation to avoid catastrophic forgetting. Therefore, it's important to precisely identify these parameters.

As shown in Tab. \ref{table:dynamic_test}, the absence of physiology-related parameters freezing (w/o $\mathcal{I_P}$) makes PhysRAP show obvious instability. At this situation, PhysRAP achieves a satisfactory MAE on PURE$^+$ (0.52 bpm), but the MAE on BUAA-MIHR$^+$ (4.39 bpm) decreases significantly. To further verify our design, we conduct HR estimation with two different physiology-related parameters identification strategies (i.e., calculation of $\mathcal{I_P}$), as shown in Tab. \ref{table:alb-physiology-related}. It's clear that the random selection strategy mistakenly optimizes the physiology-related parameters, leading to error accumulation and dissatisfied results. We further validate the effectiveness of the correlation calculation (Eq. \ref{eq:ip_expand}). As shown in row 2 of Tab. \ref{table:alb-physiology-related}, considering only the importance (diagonal elements) leads to inaccurate $\mathcal{I_P}$ estimation, thereby yielding sub-optimal results in terms of MAE (1.65 bpm) and RMSE (3.46 bpm).

\subsubsection{Impact of Future Domain Pre-adaptation} 
\begin{figure}[tb]
\centering
\includegraphics[width=1\columnwidth]{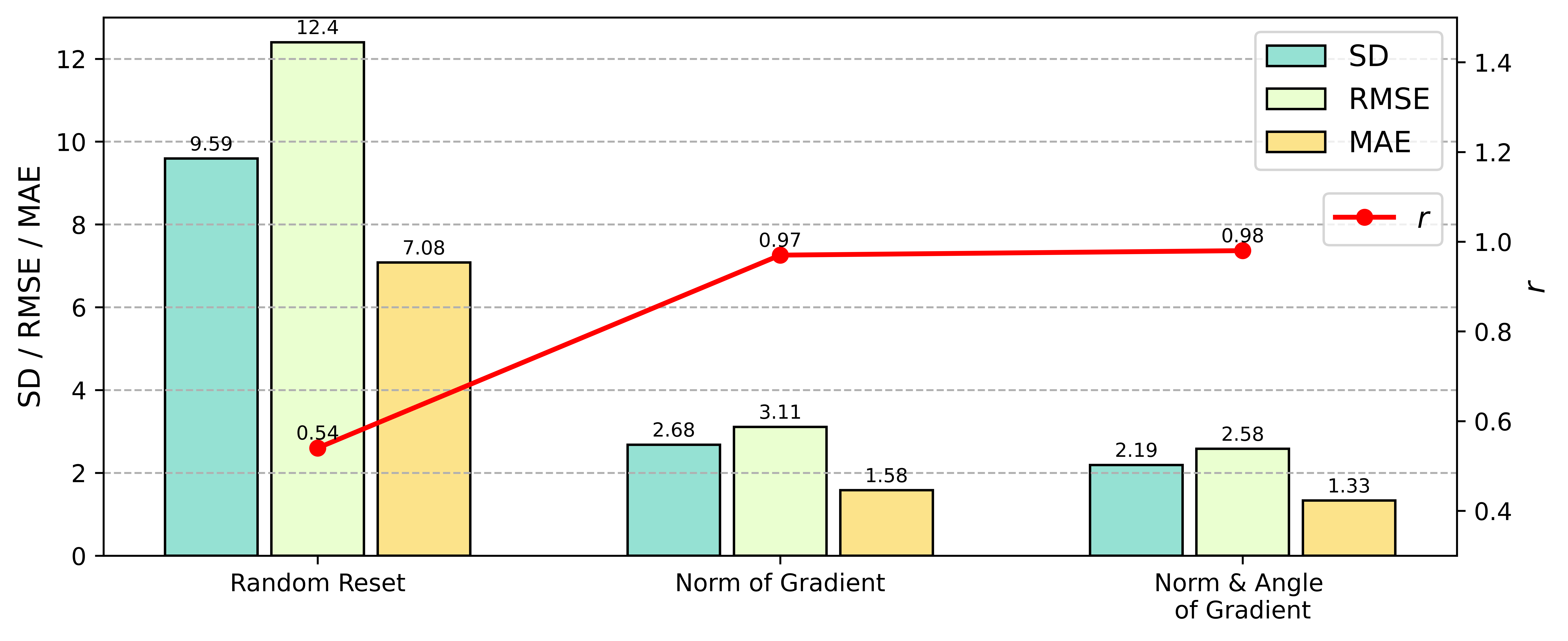}\vspace{-1em}
\caption{Visualization of the impact of preemptive gradient modification strategy.}
\label{fig:ablation_CQP_S1}
\Description{..}\vspace{-1em}
\end{figure}

We design a preemptive gradient modification strategy to prevent the model from over-adaptation to the current domain by pre-adapting to a potential future domain. As shown in Tab. \ref{table:dynamic_test}, to verify the effectiveness of this strategy (i.e., Eq. \ref{eq:modify_gc} and \ref{eq:get_w_i}), we remove this design in PhysRAP (w/o $\mathcal{F}$), which means PhysRAP will rely solely on $\mathcal{G}_C$ for adaptation. It's clear that after removing this strategy, PhysRAP exhibits an increasing performance degradation over time, which proves the effectiveness of the proposed preemptive gradient modification.

Furthermore, we investigate different variants of the gradient modification strategy, and the corresponding ablation experiments are shown in Fig. \ref{fig:ablation_CQP_S1}. Firstly, we remove the consideration of $\langle\mathcal G_F^i,\mathcal G_C^i\rangle$, which means Eq. \ref{eq:get_w_i} becomes $w_i=1/({1+\exp(-\|\mathcal G_F^i\|/\|\mathcal G_C^i\|})$. We denote this setting as the ``Norm of Gradient''. It can be seen that this setting violates the update principle described in Sec. \ref{subsec:over-adaptation}, thereby yielding sub-optimal MAE (1.58 bpm) and RMSE (3.11 bpm). Subsequently, similar to dropout \cite{dropout2014}, we restrict the model's adaptation by randomly resetting the gradients to zero, denoted as the ``Random Reset''. As shown in Fig. \ref{fig:ablation_CQP_S1}, this strategy results in a significant performance degradation (MAE = 7.08 bpm), which may be due to the loss of critical gradients.

\subsubsection{Impact of Hyper-parameters} 

\paragraph{Parameters frozen ratio}

The proportion of frozen parameters (i.e., $r_1\%$ and $r_2\%$) is the key parameter for PhysRAP to balance adaptability and memory retention. We find that the model achieved optimal results when the proportion of important parameters $r_1\%=80\%$ and the proportion of correlated parameters $r_2\%=20\%$ , as shown in Tab. \ref{table:hyper-param-a}.
\paragraph{Number of augmentations}

The number of samples in the augmented domain $N$ is crucial for the model's pseudo-label quality. As shown in Tab. \ref{table:hyper-param-b}, PhysRAP achieves the best results when $N=10$. Fewer samples lead to performance degradation, while more augmented samples do not produce significant improvements.

\paragraph{Number of samples in the future domain}

Similarly, the number of samples in the future domain $K$ determines the accuracy of gradient modification $\mathcal{F}$. As shown in Tab. \ref{table:hyper-param-b}, the best prediction performance can be achieved when $K=4$.

\paragraph{Learning rate and momentum factor}

The learning rate $\eta$ and momentum factor $\alpha$ are used to update the student network $\theta_s$ and teacher network $\theta_t$, respectively. Both excessively fast and slow updates can lead to performance degradation. To determine the value of them, we conduct ablation studies and find that the best value $\eta=1e-4,\alpha=0.99$, as shown in Tab. \ref{table:hyper-param-c}.

\section{Conclusion}
\label{sec:conclusion}

In summary, this work introduces a novel framework for rPPG measurement, namely PhysRAP, which aims to address the dynamic domain shift problem in deployment scenarios. Before adapting to inference environment, PhysRAP evaluates the model's uncertainty in the current domain, thereby identifying physiology-related knowledge and isolating it to eliminate catastrophic forgetting. Moreover, updating on the current domain may lead to over-adaptation, which hampers the model's ability to adapt to future domains. PhysRAP proactively adapts to potential future domains, thereby preventing over-adaptation. Extensive experiments demonstrate that our method attains state-of-the-art performance, particularly in handling continual domain shifts.

\bibliographystyle{ACM-Reference-Format}
\bibliography{sample-base}
\appendix
\section{Further Analysis}
\subsection{Calculation of the Non-diagonal Elements in Weight-level FIM}

As discussed in the main text, some pioneers \cite{WangCoTTA2022,BrahmaPETAL2023,CuiHoCoTTA2024} measure the importance of parameters using the parameter-level Fisher information matrix (FIM) $\mathbf{F}_{param}$. However, the larger number of parameters makes the computation of $\mathbf{F}_{param}$ extremely challenging. To address this limitation, we propose the weight-level FIM  $\mathbf{F}_{weight}$, which treats parameters belonging to the same weight (e.g., the weights and biases of a convolutional kernel) as a unit and calculates the importance of this unit. $\mathbf{F}_{weight}$ could be obtained by:
\begin{equation}
\label{eq:get_f}
\begin{split}
	\mathbf{F}_{weight}^{i,j}=\begin{cases}
    \frac{1}{P_{w_i}P_{w_j}}\sum_{l=1}^{P_{w_i}}\sum_{k=1}^{P_{w_j}}\left (\frac{\partial  \mathcal{U}_{\mathcal{T}_t}}{\partial \theta_s^i}\right )\left (\frac{\partial  \mathcal{U}_{\mathcal{T}_t}}{\partial \theta_s^j}\right )^\top_{kl}\ \mathrm{if}\ i\neq j\\
	\frac{1}{P_{w_i}}\sum_{k=1}^{P_{w_i}}\left (\frac{\partial  \mathcal{U}_{\mathcal{T}_t}}{\partial \theta_s^i}\right )^2_k\ \mathrm{otherwise},	
	\end{cases}
\end{split}
\end{equation}
where $P_{w_i}$ denotes the number of parameters in the $i$-th weight $\theta_s^i$ and $P=\sum_{i=1}^MP_{w_i}$. Although $\mathbf{F}_{weight}$ supports the comprehensive consideration of the importance and correlation of weights, the computation of its non-diagonal elements still involves an excessive number of multiplications. To simplify the computation when $i\neq j$, we can rewrite the formula as follows:
\begin{equation}
    \begin{split}
        \mathbf{F}_{weight}^{i,j}
        &=\frac{1}{P_{w_i}P_{w_j}}\sum_{l=1}^{P_{w_i}}\sum_{k=1}^{P_{w_j}}\left (\frac{\partial  \mathcal{U}_{\mathcal{T}_t}}{\partial \theta_s^i}\right )\left (\frac{\partial  \mathcal{U}_{\mathcal{T}_t}}{\partial \theta_s^j}\right )^\top_{kl}
        \\&=\frac{1}{P_{w_i}P_{w_j}}\sum_{l=1}^{P_{w_i}}\sum_{k=1}^{P_{w_j}}\left (\frac{\partial  \mathcal{U}_{\mathcal{T}_t}}{\partial \theta_s^i}\right )_l\left (\frac{\partial  \mathcal{U}_{\mathcal{T}_t}}{\partial \theta_s^j}\right )_k
        \\&=\left(\frac{1}{P_{w_i}}\sum_{l=1}^{P_{w_i}}\left (\frac{\partial  \mathcal{U}_{\mathcal{T}_t}}{\partial \theta_s^i}\right )_l\right)\left(\frac{1}{P_{w_j}}\sum_{k=1}^{P_{w_j}}\left (\frac{\partial  \mathcal{U}_{\mathcal{T}_t}}{\partial \theta_s^j}\right )_{k}\right)
        \\&=\bar{\boldsymbol{g}}_i\cdot \bar{\boldsymbol{g}}_j,
    \end{split}
\end{equation}
where $\bar{\boldsymbol{g}}_i$ and $\bar{\boldsymbol{g}}_j$ are the average gradients with respect to the weight units $\theta_s^i$ and $\theta_s^j$, respectively, defined as:
\begin{equation}
    \bar{\boldsymbol{g}}_i = \frac{1}{P_{w_i}} \sum_{l=1}^{P_{w_i}} \left( \frac{\partial \mathcal{U}_{\mathcal{T}_t}}{\partial \theta_s^i} \right)_l,\ \bar{\boldsymbol{g}}_j = \frac{1}{P_{w_j}} \sum_{k=1}^{P_{w_j}} \left( \frac{\partial \mathcal{U}_{\mathcal{T}_t}}{\partial \theta_s^j} \right)_k.
\end{equation}

By first calculating $\bar{\boldsymbol{g}}_i$ and $\bar{\boldsymbol{g}}_j$ before computing $\mathbf{F}_{weight}^{i,j}$, we successfully optimize the algorithmic complexity from $\mathcal{O}(P_{w_i}\times P_{w_j})$ to $\mathcal{O}(\max(P_{w_i},P_{w_j}))$, thereby accelerating the computation in an equivalent manner.

\subsection{Analysis of the Validity of the Correction Coefficient Formula}

In the main text, we design a preemptive gradient modification strategy that optimizes adaptation to the current domain by gradients derived from the potential future domain, thereby avoiding over-adaptation and retaining the ability to adapt to future domains. Specifically, for the $i$-th weight, the strategy takes into account the gradients $\mathcal{G}_C^i$ and $\mathcal{G}_F^i$ obtained from the current domain and the future domain, respectively, and performs gradient modification using the following formula:
\begin{equation}
    \label{eq:modify_gc}
    \mathcal{F}(\mathcal{G}_C^i,\mathcal{G}_F^i)=\begin{cases}
        0\ \ \mathrm{if}\ \cos\langle\mathcal G_F^i,\mathcal G_C^i \rangle <0\\
        w_i\times\mathcal{G}_C^i\ \ \mathrm{otherwise,}
    \end{cases}
\end{equation}
where $w_i\in(0,1)$ denotes the correction coefficient for the current gradient, which could be calculated by:
\begin{equation}
    \label{eq:get_w_i}
    \begin{split}
        w_i=&\frac{1}{1+\exp\left(-\frac{\|\mathcal G_F^i\|}{\|\mathcal G_C^i\|}\left(\cos\langle\mathcal G_F^i,\mathcal G_C^i \rangle-\frac{\sqrt 2}{2}\right) \right)}.
    \end{split}
\end{equation}

In the main text, we summarize the three properties of $w_i$: (i) When the angle is small, the directions of the two gradients are considered to be close, and a large step size update can be performed; (ii) When the angle is medium, there is some conflict between the directions of the two gradients, and the step size of the update decreases as the magnitude of $\mathcal{G}_F^i$ increases; (iii) When the angle is large, updates to $\mathcal{G}_C^i$ should be stopped. Here, we will verify these properties from a mathematical perspective.

Firstly, we simplify the Eq. \ref{eq:get_w_i} by using \( x_i \) and \( y_i \) to represent the norm ratio \( \|\mathcal G_F^i\|/\|\mathcal G_C^i\| \) and the cosine of the angle \( \cos\langle\mathcal G_F^i,\mathcal G_C^i \rangle \), respectively. This allows us to analyze the function \( w_i=f(x_i, y_i) \) as a binary function and study its properties through partial derivatives.

In this case, Eq. \ref{eq:get_w_i} can be rewritten as:
\begin{equation}
    w_i=f(x_i, y_i) = \frac{1}{1 + \exp\left(-x_i \left( y_i-\frac{\sqrt{2}}{2} \right)\right)},
\end{equation}
where \( x =  \|\mathcal G_F^i\|/\|\mathcal G_C^i\| \) is the norm ratio, and \( y = \cos\langle\mathcal G_F^i,\mathcal G_C^i \rangle \) denotes the cosine of the angle between the two vectors.

Afterward, to analyze the properties of \(w_i= f(x_i, y_i) \), we compute the partial derivatives with respect to \( x_i \) and \( y_i \):
\begin{equation}
    \frac{\partial w_i}{\partial x_i} = \frac{\partial}{\partial x_i} \left( \frac{1}{1 + \exp\left(-x_i \left( y_i-\frac{\sqrt{2}}{2} \right)\right)} \right).
\end{equation}
Let \( z_i = -x_i \left( y_i-\frac{\sqrt{2}}{2} \right) \), then:
\begin{equation}
    w_i = \frac{1}{1 + \exp(z_i)},
\end{equation}
The derivative of \( w_i \) with respect to \( z_i \) is:
\begin{equation}
    \frac{\partial w_i}{\partial z_i} = -\frac{\exp(z_i)}{(1 + \exp(z_i))^2},
\end{equation}
Thus:
\begin{equation}
    \frac{\partial w_i}{\partial x_i} = \frac{\partial w_i}{\partial z_i} \cdot \frac{\partial z_i}{\partial x_i} = -\frac{\exp(z_i)}{(1 + \exp(z_i))^2} \cdot \left(-\left( y_i-\frac{\sqrt{2}}{2} \right)\right),
\end{equation}
Simplify this formula, we can obtain:
\begin{equation}
\label{eq:w/x}
    \frac{\partial w_i}{\partial x_i} = \frac{\exp(z_i) \left( y_i-\frac{\sqrt{2}}{2} \right)}{(1 + \exp(z_i))^2},
\end{equation}
where \( z_i = -x_i \left( y_i-\frac{\sqrt{2}}{2} \right)\). Similarly, we can calculate the partial derivative with respect to $y_i$:
\begin{equation}
\label{eq:w/y}
    \frac{\partial w_i}{\partial y_i} = -\frac{x_i \exp(z_i)}{(1 + \exp(z_i))^2}.
\end{equation}

Based on the formulation of Eq. \ref{eq:w/x} and \ref{eq:w/y}, we could summarize the function properties:
\begin{itemize}
    \item When \(  y_i-\frac{\sqrt{2}}{2} > 0 \) (i.e., $\langle\mathcal G_F^i,\mathcal G_C^i \rangle<\pi/4$), \( \frac{\partial w_i}{\partial x_i} > 0 \), indicating that \( w_i \) increases with \( x_i \).
    \item When \(  y_i-\frac{\sqrt{2}}{2} < 0 \), \( \frac{\partial w_i}{\partial x_i} < 0 \) (i.e., $\langle\mathcal G_F^i,\mathcal G_C^i \rangle>\pi/4$), indicating that \( w_i \) decreases with \( x_i \).
    \item \( \frac{\partial w_i}{\partial y_i} < 0 \), indicating that \( w_i \) decreases as \( y_i \) increases.
\end{itemize}

\section{Further Experiments}

\subsection{Inference Time of PhysRAP}

\begin{table}[tb]
\centering
\caption{HR estimation results and inference times (ms) under CTTA protocol. The symbols $\triangle$, $\star$, and $\ddagger$ denote the traditional, deep learning-based, and TTA methods (based on ResNet3D-18 \cite{HaraResNet3D2018}), respectively. Best results are marked in \textbf{bold} and second best in \underline{underline}.}
\begin{tabular}{lccccc}
\toprule
\multirow{2}{*}{Method} &\multicolumn{4}{c}{MEAN}&\multirow{2}{*}{Time$\downarrow$} \\
\cmidrule(lr){2-5}
& SD$\downarrow$ & MAE$\downarrow$ & RMSE$\downarrow$ & $r\uparrow$&\\
\midrule
GREEN$^\triangle$\cite{VerkruysseGreen2008} & 15.8 & 37.2 & 41.3 & 0.11 & \underline{20}\\
ICA$^\triangle$\cite{PohICA2010} & 10.5 & 10.9 & 14.4 & 0.78 & 67\\
POS$^\triangle$\cite{WangPOS2017} & 8.87 & 7.49 & 10.2 & 0.82 & 71\\
\midrule
PhysNet$^\star$\cite{YuPhysNet2019} & 13.2 & 8.48 & 15.2 & 0.57 & \textbf{14}\\
PhysMamba$^\star$\cite{LuoPhysMamba2024} & 13.5 & 8.76 & 16.1 & 0.61 &55\\
PhysFormer$^\star$\cite{YuPhysFormer2022} & 8.39 & 4.56 & 8.82 & 0.85 & 29\\
RhythmMamba$^\star$\cite{zou2024rhythmmamba} & \underline{4.46} & 3.10 & \underline{4.83} & \underline{0.93} & 42\\
\midrule
CoTTA$^{\ddagger\star}$\cite{WangCoTTA2022} & 8.90 & 6.96 & 11.0 & 0.67 & 29\\
DA-TTA$^{\ddagger\star}$\cite{WangDaTTA2024} & 7.53 & 3.23 & 7.90 & 0.91 & 51\\
RoTTA$^{\ddagger\star}$\cite{YuanRoTTA2023} & 7.88 & 4.16 & 8.29 & 0.86 & 30\\
PETAL$^{\ddagger\star}$\cite{BrahmaPETAL2023} & 5.69 & \underline{2.70} & 6.09 & \underline{0.93} & 95\\
\textbf{PhysRAP}$^{\ddagger\star}$ & \textbf{2.19 }& \textbf{1.33} & \textbf{2.58} & \textbf{0.98} & 48\\
\bottomrule
\end{tabular}
\label{table:inf-time}
\end{table}

In the real-world deployment of CTTA rPPG models, inference speed is also a core factor of the model. To verify the inference speed of PhysRAP, we present the inference speed (milliseconds per frame) of various methods in Tab. \ref{table:inf-time}. The inference time per frame is calculated with the video input size 3$\times$300$\times$128$\times$128  ($C\times T\times H\times W$) on a single RTX 3090 GPU for all frameworks. It can be seen that PhysRAP achieves the best performance without incurring significant additional inference time. PhysRAP can infer approximately 21 frames per second, which is fully capable of supporting real-time rPPG measurement.

\subsection{Single Domain Testing Results}

\begin{table}[tb]
\centering
\caption{Single domain HR estimation results on VIPL-HR. Here, the baseline denotes the ResNet3D-18 \cite{HaraResNet3D2018}.}
\begin{tabular}{lcccc}
\toprule
Method & SD$\downarrow$ & MAE$\downarrow$ & RMSE$\downarrow$ & $r\uparrow$ \\
\midrule
SAMC$^\triangle$\cite{TulyakovSAMC2016} & 18.0 & 15.9 & 21.0 & 0.11 \\
CHROM$^\triangle$\cite{DeCHROM2013} & 15.1 & 11.4 & 16.9 & 0.28 \\
POS$^\triangle$\cite{WangPOS2017} & 15.3 & 11.5 & 17.2 & 0.30 \\
\midrule
PhysNet$^\star$\cite{YuPhysNet2019} & 14.9 & 10.8 & 14.8 & 0.20 \\
CVD$^\star$\cite{NiuCVD2020} & 7.92 & 5.02 & 7.97 & 0.79 \\
PhysFormer$^\star$\cite{YuPhysFormer2022} & 7.74 & 4.97 & 7.79 & 0.78  \\
NEST$^\star$\cite{LuNEST2023} & 7.49 & 4.76 & 7.51 & \underline{0.84} \\
DOHA$^\star$\cite{DOHASun2023} & - & 4.87 & 7.64 & 0.83 \\
rPPG-HiBa$^\star$\cite{HiBaLu2024} &\underline{7.26} & \underline{4.47} & \underline{7.28} & \textbf{0.85}\\
\midrule
Baseline$^{\ddagger\star}$ & 9.29 & 5.56 & 9.41 & 0.62 \\
Baseline+\textbf{ours}$^{\ddagger\star}$ & 7.47 & 4.78 & 7.67 & 0.75 \\
PhysFormer+\textbf{ours}$^{\ddagger\star}$ & \textbf{6.96} &\textbf{ 4.12} & \textbf{6.97} & \underline{0.84} \\
\bottomrule
\end{tabular}
\label{table:single_test}
\end{table}

Here, we simplify the CTTA protocol to the TTA protocol, where the model still faces the issue of distribution shift between training and testing data, but adaptation is required only on a single domain. According to \cite{LuNEST2023}, VIPL-HR dataset has multiple complex scenes and recording devices and cannot be considered as a single domain. Therefore, we test PhysRAP with different baselines using the 5-fold cross-validation protocol \cite{LuNEST2023,YuPhysFormer2022} on VIPL-HR. As shown in Tab. \ref{table:single_test}, we first report the HR estimation results of the baseline model (i.e., ResNet3D-18 \cite{HaraResNet3D2018}) and PhysRAP based on this baseline. Clearly, our PhysRAP framework achieves a significant improvement of 0.78 bpm in MAE (compared to 5.56 bpm) and an improvement of 1.74 bpm in RMSE (compared to 9.41 bpm). Furthermore, when we employ an end-to-end rPPG model as the baseline (i.e., PhysFormer \cite{YuPhysFormer2022}), we observe that PhysRAP, implemented based on this baseline, achieves the best results in terms of SD, MAE, and RMSE.

\subsection{More Details of the Ablation Study}

\begin{table*}[t]
\centering
\caption{The Details of Ablation Studies under CTTA protocol. The default settings of PhysRAP and their corresponding results are highlighted with \colorbox{gray!20}{shading}. Here, D., P., W., N., A., and G. are short for diagonal, parameter, weight, norm, angle, and gradients, respectively.}
\setlength{\tabcolsep}{0.37mm}{
\begin{tabular}{l|ccc|ccc|ccc|ccc|ccc|ccc|ccc}
\toprule
Time & \multicolumn{18}{l|}{$\ t\xrightarrow{\hspace*{12.8cm}}$}& \\ \hline
\multirow{2}{*}{Components} & \multicolumn{3}{c}{UBFC-rPPG\cite{BobbiaUBFC2019}} & \multicolumn{3}{c}{UBFC-rPPG$^+$} & \multicolumn{3}{c}{PURE\cite{StrickerPURE2014}} &\multicolumn{3}{c}{PURE$^+$} & \multicolumn{3}{c}{BUAA-MIHR\cite{XiBUAA2020}} & \multicolumn{3}{c}{BUAA-MIHR$^+$} & \multicolumn{3}{c}{MEAN}\\
\cmidrule(lr){2-4} \cmidrule(lr){5-7} \cmidrule(lr){8-10} \cmidrule(lr){11-13} \cmidrule(lr){14-16} \cmidrule(lr){17-19} \cmidrule(lr){20-22}
& {MAE} & RMSE & $r$& MAE & RMSE & $r$& MAE & RMSE & $r$& MAE & RMSE & $r$& MAE & RMSE & $r$& MAE & RMSE & $r$& MAE & RMSE & $r$\\
\midrule
$r_1,r_2=70,20$&0.75&1.81&0.99&0.84&2.13&0.99&0.92&3.52&0.98&0.29&0.73&0.99&2.63&4.01&0.95&3.35&5.42&0.90&1.46&2.93&0.97\\
$r_1,r_2=80,10$&0.73&1.98&0.99&0.82&2.07&0.99&0.72&1.87&0.99&0.31&0.74&0.99&2.62&4.17&0.94&4.42&7.53&0.80&1.60&3.06&0.95\\
\rowcolor{gray!20}$r_1,r_2=80,20$& {{0.81}} & {{1.84}} & {{0.99}}& {{0.85}} & {{2.13}} & {{0.99}}& {{1.10}} & {3.78} & {0.99} & {{0.31}} & {{0.75}} & {{1.00}}& {{2.46}} & {{3.33}} & {{0.98}}& {{2.48}} & {{3.65}}& {0.96} & {{1.33}} & {{2.58}} &{0.98}\\
$r_1,r_2=80,30$&0.72&1.98&0.99&0.82&2.06&0.99&0.62&1.57&0.99&0.30&0.73&0.99&2.20&2.61&0.99&5.39&9.15&0.69&1.67&3.01&0.94\\
$r_1,r_2=90,20$&0.72&1.98&0.99&0.82&2.06&0.99&0.62&1.58&0.99&0.28&0.74&0.99&3.06&4.58&0.30&4.23&5.93&0.12&1.62&2.81&0.73\\
\midrule
$N,K=5,4$&0.81&1.84&0.99&0.85&2.14&0.99&1.10&3.79&0.98&0.29&0.75&0.99&5.88&11.1&0.52&7.75&13.5&0.29&2.78&5.55&0.80\\
$N,K=10,2$&0.88&2.29&0.99&0.86&2.14&0.99&1.00&3.25&0.99&0.30&0.75&0.99&6.62&12.4&0.40&8.74&14.7&0.21&3.06&5.94&0.76\\
\rowcolor{gray!20}$N,K=10,4$& {{0.81}} & {{1.84}} & {{0.99}}& {{0.85}} & {{2.13}} & {{0.99}}& {{1.10}} & {3.78} & {0.99} & {{0.31}} & {{0.75}} & {{1.00}}& {{2.46}} & {{3.33}} & {{0.98}}& {{2.48}} & {{3.65}}& {0.96} & {{1.33}} & {{2.58}} &{0.98}\\
$N,K=10,8$&0.78&1.83&0.99&0.85&2.14&0.99&1.44&4.56&0.98&0.49&1.26&0.99&3.45&2.80&0.88&3.56&5.57&0.90&1.76&3.03&0.96\\
$N,K=15,4$&0.81&1.84&0.99&0.85&2.14&0.99&0.85&2.74&0.99&0.30&0.74&0.99&2.44&4.17&0.97&3.09&4.13&0.88&1.39&2.63&0.97\\
\midrule
$\eta,\alpha=5$e-5, 0.99&0.95&2.34&0.98&0.93&2.32&0.98&1.23&4.13&0.98&0.38&0.93&0.99&2.76&3.89&0.96&2.95&4.57&0.93&1.54&3.03&0.98\\
$\eta,\alpha=1$e-4, 0.985&0.81&1.84&0.99&0.84&2.13&0.99&0.84&2.72&0.99&0.30&0.75&0.99&10.0&16.0&0.16&15.7&20.6&0.09&4.57&7.35&0.68\\
\rowcolor{gray!20}$\eta,\alpha=1$e-4, 0.99& {{0.81}} & {{1.84}} & {{0.99}}& {{0.85}} & {{2.13}} & {{0.99}}& {{1.10}} & {3.78} & {0.99} & {{0.31}} & {{0.75}} & {{1.00}}& {{2.46}} & {{3.33}} & {{0.98}}& {{2.48}} & {{3.65}}& {0.96} & {{1.33}} & {{2.58}} &{0.98}\\
$\eta,\alpha=1$e-4, 0.995&0.81&1.84&0.99&0.85&2.14&0.99&1.03&3.66&0.98&0.31&0.75&0.99&2.53&3.70&0.96&2.66&4.09&0.95&1.36&2.69&0.98\\
$\eta,\alpha=2$e-4, 0.99&0.74&1.99&0.99&0.71&1.53&0.99&1.15&4.39&0.98&0.33&0.81&0.99&5.11&10.1&0.58&6.77&12.8&0.33&2.47&5.27&0.80\\
\midrule
Random Select&2.01&3.38&0.99&1.92&3.32&0.98&3.46&4.73&0.97&1.34&2.80&0.98&3.41&5.09&0.95&3.51&4.52&0.95&2.61&3.97&0.97\\
D. of W. FIM&0.80&1.84&0.99&0.85&2.13&0.99&0.90&2.74&0.99&0.77&3.85&0.98&2.92&4.36&0.94&3.68&5.85&0.89&1.65&3.46&0.96\\
\rowcolor{gray!20}W. FIM& {{0.81}} & {{1.84}} & {{0.99}}& {{0.85}} & {{2.13}} & {{0.99}}& {{1.10}} & {3.78} & {0.99} & {{0.31}} & {{0.75}} & {{1.00}}& {{2.46}} & {{3.33}} & {{0.98}}& {{2.48}} & {{3.65}}& {0.96} & {{1.33}} & {{2.58}} &{0.98}\\
\midrule
Random Reset&0.80&1.84&0.99&0.84&2.13&0.99&1.00&3.29&0.99&0.68&3.17&0.99&19.2&26.4&0.02&35.0&37.8&0.04&9.59&12.4&0.67\\
N. of G.&0.80&1.84&0.99&0.84&2.13&0.99&1.35&4.36&0.98&0.41&1.06&0.99&3.38&5.19&0.92&2.69&4.07&0.95&1.57&3.10&0.97\\
\rowcolor{gray!20}N. \& A. of G. & {{0.81}} & {{1.84}} & {{0.99}}& {{0.85}} & {{2.13}} & {{0.99}}& {{1.10}} & {3.78} & {0.99} & {{0.31}} & {{0.75}} & {{1.00}}& {{2.46}} & {{3.33}} & {{0.98}}& {{2.48}} & {{3.65}}& {0.96} & {{1.33}} & {{2.58}} &{0.98}\\
\bottomrule
\end{tabular}}
\label{table:dynamic_test}
\end{table*}

To avoid confusion, we only present the $\mathrm{MEAN}$ metric in the ablation experiments in the main text. However, the specific details of these experiments under the CTTA protocol also reflect their performance in avoiding catastrophic forgetting and over-adaptation. Therefore, we present the detailed specifics of all ablation experiments in Tab. \ref{table:dynamic_test}, aiming to provide a new perspective for analyzing the effectiveness of the key components of PhysRAP.

\subsubsection{Impact of Parameters Frozen Ratio} 
As discussed in the main text, \( r_1\% \) and \( r_2\% \) jointly determine the proportion of physiologically relevant parameters to be frozen, and only with appropriate values (i.e., $r_1\%=80\%,r_2\%=20\%$) can the global optimal solution be precisely achieved.

\subsubsection{Impact of Number of Augmentations} 
The number of augmentations \( N \) determines the accuracy of the model's estimation of pseudo-labels. Therefore, as \( N \) increases, the model performance initially improves and then plateaus. Meanwhile, too few augmentations (\( N = 5,K=4 \)) may affect the precision of simulating the potential future domain, thereby causing the model to exhibit a certain degree of over-adaptation risk (manifested by deteriorated performance on the last two domains).

\subsubsection{Impact of Number of Samples in the Future Domain} 
PhysRAP is sensitive to the number of samples in the future domain $K$. An insufficient number of samples ($N=10,K=2$) may cause fluctuations in the potential future domain, thereby severely affecting the model's adaptation to the actual future domain.

\subsubsection{Impact of Learning Rate and Momentum Factor} 
PhysRAP requires an appropriate learning rate and momentum factor. Too slow adaptation may slow down the model's convergence speed, while too fast adaptation may prevent the model from finding the optimal solution. Both scenarios can lead to suboptimal results. In particular, when the learning rate is too high ($\eta=2$e-$4,\alpha=0.99$), the model may deviate from the optimizable region during adaptation, manifesting as a gradual loss of adaptability.

\subsubsection{Impact of Physiology-related Parameters Freezing} 
As shown in Tab. \ref{table:dynamic_test}, our design primarily demonstrates the model's ability to maintain long-term adaptability, which stems from PhysRAP's enhanced capability to accurately identify physiologically relevant parameters.

\subsubsection{Impact of Future Domain Pre-adaptation} 
From the perspective of long-term adaptability, the future domain pre-adaptation we proposed effectively alleviates the over-adaptation problem, allowing the model to retain its adaptability even after adapting to multiple domains.

\section{Specific Network Architecture}

\begin{table*}[t]
\caption{Parameter illustration of network architectures. C3d denotes the 3D convolutional layer, BN represents the batch normalization, [+$\mathbf{F}$] denotes the residual connection, and $\mathrm{DS}\downarrow_x$ denotes down-sampling with the scale of $x$.}
\centering
\renewcommand\arraystretch{1.3}
\setlength{\tabcolsep}{1.7mm}{
\begin{tabularx}{1\textwidth}{c|l|l}
\toprule
Module & Input  $\rightarrow$  Output & Layer Operation \\
\midrule

&$\mathbf{V}(3,T,H,W) \rightarrow \mathbf{F}_0(D,T,\frac{H}{2},\frac{W}{2})$ & C3d[5,2,2] $\to$ BN $\to$ ReLU $\to$ MaxPool\\
&$\mathbf{F}_0(D,T,\frac{H}{2},\frac{W}{2}) \rightarrow \mathbf{F}_0'(D,T,\frac{H}{2},\frac{W}{2})$ & C3d[3,1,1] $\to$ BN $\to$ ReLU $\to$ C3d[3,1,1] $\to$ BN [+$\mathbf{F_0}$] $\to$ ReLU\\
&$\mathbf{F}_0'(D,T,\frac{H}{2},\frac{W}{2}) \rightarrow \mathbf{F}_1(D,T,\frac{H}{2},\frac{W}{2})$& C3d[3,1,1] $\to$ BN $\to$ ReLU $\to$ C3d[3,1,1] $\to$ BN [+$\mathbf{F_0'}$] $\to$ ReLU\\
&$\mathbf{F}_1(D,T,\frac{H}{2},\frac{W}{2}) \rightarrow \mathbf{F}_1'(2D,\frac{T}{2},\frac{H}{4},\frac{W}{4})$& C3d[3,2,1] $\to$ BN $\to$ ReLU $\to$ C3d[3,1,1] $\to$ BN [+$\mathrm{DS}\downarrow_2(\mathbf{F_1})$] $\to$ ReLU\\
&$\mathbf{F}_1'(2D,\frac{T}{2},\frac{H}{4},\frac{W}{4}) \rightarrow\mathbf{F}_2(2D,\frac{T}{2},\frac{H}{4},\frac{W}{4})$&  C3d[3,1,1] $\to$ BN $\to$ ReLU $\to$ C3d[3,1,1] $\to$ BN [+$\mathbf{F_1'}$] $\to$ ReLU\\
ResNet3D-18&$\mathbf{F}_2(2D,\frac{T}{2},\frac{H}{4},\frac{W}{4}) \rightarrow\mathbf{F}_2'(4D,\frac{T}{4},\frac{H}{8},\frac{W}{8})$&  C3d[3,2,1] $\to$ BN $\to$ ReLU $\to$ C3d[3,1,1] $\to$ BN [+$\mathrm{DS}\downarrow_2(\mathbf{F_2})$] $\to$ ReLU\\
&$\mathbf{F}_2'(4D,\frac{T}{4},\frac{H}{8},\frac{W}{8}) \rightarrow\mathbf{F}_3(4D,\frac{T}{4},\frac{H}{8},\frac{W}{8})$&  C3d[3,1,1] $\to$ BN $\to$ ReLU $\to$ C3d[3,1,1] $\to$ BN [+$\mathbf{F_2'}$] $\to$ ReLU\\
&$\mathbf{F}_3(4D,\frac{T}{4},\frac{H}{8},\frac{W}{8}) \rightarrow\mathbf{F}_3'(4D,\frac{T}{4},\frac{H}{8},\frac{W}{8})$&  C3d[3,1,1] $\to$ BN $\to$ ReLU $\to$ C3d[3,1,1] $\to$ BN [+$\mathbf{F_3}$] $\to$ ReLU\\
&$\mathbf{F}_3'(4D,\frac{T}{4},\frac{H}{8},\frac{W}{8}) \rightarrow\mathbf{F}_4(4D,\frac{T}{4},\frac{H}{8},\frac{W}{8})$&  C3d[3,1,1] $\to$ BN $\to$ ReLU $\to$ C3d[3,1,1] $\to$ BN [+$\mathbf{F_3'}$] $\to$ ReLU\\
&$\mathbf{F}_4(4D,\frac{T}{4},\frac{H}{8},\frac{W}{8}) \rightarrow\mathbf{F}_{fin}(4D,T,\frac{H}{8},\frac{W}{8})$&  C3d$^\top$[4.1.1,2.1.1,1.0.0] $\to$ BN $\to$ ELU $\to$ C3d$^\top$[4.1.1,2.1.1,1.0.0] $\to$ BN $\to$ ELU\\
&$\mathbf{F}_{fin}(4D,T,\frac{H}{8},\frac{W}{8}) \rightarrow \boldsymbol{s}_{pre}(T)$&  AvgPool $\to$ C3d[1,1,1] $\to$ Squeeze\\
\bottomrule
\end{tabularx}}

\label{table:architecture}
\end{table*}

Here, we describe the implementation details of ResNet3D-18, including the specific backbone network and the structure of the rPPG prediction head.

ResNet3D-18 is an end-to-end CNN-based model, mainly comprising feature embedding, eight residual blocks for feature encoding, and several projection layers for rPPG signal estimation. Specifically, the network structure is shown in Table \ref{table:architecture}.

\section{Details of Data Augmentation Functions}

In the main text, we utilize data augmentation functions in two procedures: 1) Augmented domain $\mathcal{D_A}$ generation (i.e., the facial video augmenter $\mathcal{A}$), and 2) Dataset augmentation (i.e., PURE -> PURE$^+$). Here, we provide the specific implementation details of the data augmentation functions in these procedures:

\begin{itemize}
    \item Flipping: Horizontal flip.
    \item Gaussian Noise: Mean and variance are 0 and 0.1.
    \item Gamma Correction: Gamma factor $\sim U(0.5,1.5)$.
    \item Gaussian Blur: Kernel size is $5\times5$, sigma $\sim U(0.5,1.5)$.
    \item Cropping: Randomly select a region larger than $\frac{1}{4}$ of the original frame.
    \item Temporal Reversal: Reverse the frame sequence.
\end{itemize}

\end{document}